\documentclass[12pt,a4paper,final]{iopart}

\usepackage{iopams}
\usepackage{ifthen}
\usepackage{cite}
\expandafter\let\csname equation*\endcsname\relax
\expandafter\let\csname endequation*\endcsname\relax

\usepackage{amsmath}
\usepackage{iopams}
\usepackage{graphicx}
\usepackage[breaklinks=true,colorlinks=true,linkcolor=blue,urlcolor=blue,citecolor=blue]{hyperref}

\graphicspath{{figs/}}

\newcommand{\D}{\mathrm{d}}
\newcommand{\E}{\mathrm{e}}
\newcommand{\I}{\mathrm{i}}
\newcommand{\abs}[1]{\left\lvert{#1}\right\rvert}
\newcommand{\cc}{\mathrm{cc}}
\renewcommand{\ss}{\mathrm{ss}}
\newcommand{\cs}{\mathrm{cs}}
\renewcommand{\sc}{\mathrm{sc}}

\newcommand{\TT}{\mathcal{T}}

\begin{document}

\title[Spectral-density correlations]{Frequency-frequency correlations of single-trajectory spectral densities of Gaussian processes}

\author{Alessio Squarcini$^{1,2,3}$, Enzo Marinari$^{4,5}$, Gleb Oshanin$^6$, Luca Peliti$^7$ \& Lamberto Rondoni$^{8, 9}$}

\address{$^1$ Institut f\"ur Theoretische Physik, Universit\"at Innsbruck, 
Technikerstrasse 21A,
A-6020 Innsbruck, Austria\\
$^2$ Max-Planck-Institut f\"ur Intelligente Systeme, Heisenbergstr. 3, D-70569, Stuttgart,
Germany\\
$^3$ IV. Institut f\"ur Theoretische Physik, Universit\"at Stuttgart, Pfaffenwaldring 57,
D-70569 Stuttgart, Germany\\
$^4$ Dipartimento di Fisica, Sapienza Universit{\`a} di Roma, P.le A.
Moro 2, I-00185 Roma, Italy\\
$^5$INFN, Sezione di Roma 1 and Nanotech-CNR, UOS di Roma, P.le A.
Moro 2, I-00185 Roma, Italy\\
$^6$ Sorbonne Université, CNRS, Laboratoire de Physique Th\'eorique de la Mati\`{e}re Condens\'ee (UMR CNRS 7600), 4 place Jussieu,
75252 Paris Cedex 05, France\\
$^7$ Santa Marinella Research Institute, Santa Marinella, Italy\\
$^8$ Dipartimento di Scienze Matematiche, Politecnico di Torino, Corso Duca degli Abruzzi 24, 10129 Torino, Italy\\
$^9$ INFN, Sezione di Torino, Via P. Giuria 1, 10125 Torino, Italy}

\begin{abstract}
We investigate the stochastic behavior of the single-trajectory spectral density $S(\omega,\TT)$ of several Gaussian stochastic processes, i.e., Brownian motion, the Orn\-stein-Uhl\-en\-beck process, the Brownian gyrator model and fractional Brownian motion, as a function of the frequency~$\omega$ and the observation time~$\TT$. We evaluate in particular  the variance and the frequency-frequency correlation of~$S(\omega,\TT)$ for different values of~$\omega$. We show that these properties exhibit different behaviors for different physical cases and can therefore be used as a sensitive probe discriminating between different kinds of random motion. These results may prove quite useful in the analysis of experimental data.\end{abstract}


\maketitle


\section{Introduction}

The power spectral density (PSD) of a stochastic process contains a wealth of information 
on its temporal evolution and correlations. In particular, 
the asymptotics of the frequency dependence of the PSD of a diffusion process characterise
the short- and large-time behaviour of the process under consideration 
and can often discriminate anomalous from standard behavior
(see, e.g., \cite{norton,flandrin,eli,eli2,dechant,olivier,dean,flyv,satya,alessio,carlos}).
As a consequence, a substantial knowledge of the PSDs of rather diverse 
processes has nowadays been accumulated.   Examples contain, for instance, the spectral analysis of time-series 
 associated with the  loudness of musical recordings \cite{voss,geisel}, 
 noise in graphene devices \cite{balandin},
  evolution of climate data \cite{talkner}, 
  fluorescence intermittency in nanosystems \cite{fran}, 
  time gaps between large earthquakes \cite{sornette}, 
  current fluctuations in nanoscale electrodes \cite{krapf} and across nanopores \cite{ramin1}, 
  statistics of blinking quantum dots \cite{QD}, dynamics of tracers 
  in artificially crowded fluids \cite{mat}, as well
as the evolution of velocities of motile amoebae \cite{li,ped}, trajectories of nano-machines, e.g., of the  Brownian gyrator \cite{alberto1,alberto2}, of active Brownian motion~\cite{solon}, of anomalously diffusing walkers evolving in optical traps \cite{felix}, and of membrane proteins and other subordinated random walks \cite{eli3}.  
Other examples can be found, e.g.,  in~\cite{a}.

In most of these works, the attention is focused on the limit, for the observation time~$\TT$ going to infinity, of the average $\mu(\omega,\TT)=\overline{S(\omega,\TT)}$ of the PSD over all realizations of the process for a given (angular) frequency~$\omega$ (see, e.g., \cite{norton}). However, since $S(\omega,\TT)$ for each realization of the process is a random quantity, there is a great deal of untapped information in the statistical properties of the single-trajectory PSD. This information can be experimentally retrieved thanks to the recent progress in tracking trajectories of individual particles. One can thus be interested, e.g., in the variance of $S(\omega,\TT)$ or in the correlation of the spectral density for two different values of $\omega$ for the same realization.

These questions have been addressed in particular for the time series 
associated with blinking events in quantum dots \cite{eli}, 
in standard Brownian motion \cite{krapf1}   
and in several anomalous diffusion processes \cite{krapf2,krapf3,sposini1,sposini2},  and also
for the trajectories of the Brownian gyrator \cite{sara}. For instance, in~\cite{eli} (see also \cite{norton} for some other examples) the distribution of $S(\omega,\TT)$ was obtained for the time series of blinking events in quantum dots, and shown to be exponentially distributed in the limit $\TT \to \infty$. However, this is the only system, to our knowledge, in which the frequency-frequency correlation of the PSD has been analyzed~\cite{eli}. It is therefore interesting to investigate the behavior of the frequency-frequency correlation of the PSD in different Gaussian processes.

It was shown for several distinct examples of centered Gaussian processes,\footnote{These examples include standard Brownian motion \cite{krapf1}, fractional Brownian motion with Hurst index $H \in (0,1)$ \cite{krapf2,krapf3}, 
scaled Brownian motion \cite{sposini1},  several kinds of diffusing-diffusivity model \cite{sposini2}, as well as
the Brownian gyrator model \cite{sara}.} that the probability distribution function (PDF) of $S(\omega,\TT)$ has a universal expression that depends only on its average $\mu(\omega,\TT)$ and on its coefficient of variation $\gamma(\omega,\TT)$, defined by
\begin{equation}
\label{gamma}
\gamma(\omega,\TT)=\frac{\sqrt{\overline{S^2(\omega,\TT)}-\mu^2(\omega,\TT)}}{\mu(\omega,\TT)}.
\end{equation}
The explicit form of the probability distribution function for these processes implies that $\gamma(\omega,\TT)$ satisfies the crucial two-sided inequality
\begin{align}
\label{crucial}
1 \leq \gamma \leq \sqrt{2} \,. 
\end{align}
This has the consequence in particular that fluctuations of the PSD are generically larger than its mean value. It was shown in~\cite{crucial} that this inequality holds in general for centered Gaussian processes. 

It was also shown, for the same processes as above, that $\gamma(\omega,\TT)$ depends on the frequency and the observation time only via the product $\omega \TT$. Importantly, for fixed $\omega > 0$ and $\TT \to \infty$, the coefficient of variation approaches a constant $\omega$-independent  value, dependent only on the spread of the process. It was thus suggested \cite{krapf2} that $\gamma$ can also serve as a robust criterion of anomalous diffusion---the issue to which we will return at the end of this paper. Moreover, the probability distribution function reached in the $\TT\to\infty$ limit depends on $S(\omega)=\lim_{\TT\to\infty}S(\omega,\TT)$ and on its average $\mu(\omega)$ only in the combination
\begin{equation}
\label{prob3}
b_{\omega}=\frac{S(\omega)}{\mu(\omega)}.
\end{equation}
Therefore, in the limit of infinite observation time, the statistical properties of $S(\omega,\TT)$ are summarized by those of the random variables $b_{\omega}$ for different values of $\omega$. It is therefore natural to investigate the properties of these variables, and in particular of their mutual dependence for different values of~$\omega$. It was shown in particular that for the blinking quantum dots \cite{eli} and some other ergodic processes (see \cite{norton}) the values of $b_{\omega}$ for different values of $\omega$ are totally uncorrelated. However, these quantities can in principle fit in three possible scenarios, namely (1) they could be totally correlated, or (2) only partially correlated, or (3) fully uncorrelated, in processes of different kinds.

In this paper we determine the frequency-frequency correlation function and the Pearson correlation coefficient  of several Gaussian stochastic processes, i.e., the standard Brownian motion, the Ornstein-Uhlenbeck process, the out-of-equilibrium Brownian gyrator model and a family of anomalous diffusion processes---i.e., fractional Brownian motion with the Hurst index $H \in (0,1)$. We show that, depending on the process under consideration, in the limit $\TT \to \infty$ 
all three above-mentioned scenarios can be realised. 
Moreover, the behavior of the correlations for finite observation time $\TT$ is very rich and exhibits different properties when one of the frequencies is kept equal to zero.

In section \ref{general} we present,  for an arbitrary Gaussian process, general expressions for the frequency-frequency correlation function and for the associated Pearson correlation coefficient. In section \ref{BM} we discuss in detail the cases of standard Brownian motion and standard Ornstein-Uhlenbeck process. Section \ref{BG} is devoted to the analysis of expressions of the correlations and of the Pearson coefficient for a Brownian gyrator model, while in section \ref{FBM} we discuss the rich behaviour of the frequency-frequency correlations of a single-trajectory PSD in case of the fractional Brownian motion with arbitrary Hurst index~$H$. In section \ref{conc2} we conclude with a brief summary of our results and a discussion. 
Some details of the (rather lengthy) intermediate calculations for fractional Brownian motion are relegated to \ref{A}.

\section{General}
\label{general}
In this section we sketch the method to evaluate the frequency-frequency correlation function of the PSD for a general Gaussian process.

The single-trajectory spectral density $S(\omega,\TT)$ of the process $X(t)$ is defined by
\begin{equation}
\label{def2}
S(\omega,\TT)=\frac{1}{\TT}\abs{\int_0^\TT\D t\;\E^{\I\omega T}\,X(t)}^2.
\end{equation}
With this definition, 
the frequency-frequency correlation function of $S(\omega,\TT)$ is given by 
\begin{equation}
\begin{split}
\label{SS}
\overline{S(\omega_1,\TT) S(\omega_2,\TT)} &= \frac{1}{\TT^2} \int_0^\TT \D t_1 \int^\TT_0 \D t_2 \int^\TT_0 \D t_3 \int^\TT_0 \D t_4 \; \overline{X(t_1) X(t_2) X(t_3) X(t_4)} \\
& \times \cos\left(\omega_1(t_1 - t_2)\right) \cos\left(\omega_2 (t_3 - t_4)\right) \,.
\end{split}
\end{equation}
Concentrating from now on exclusively on Gaussian processes $X(t)$, we take advantage of Wick's theorem to obtain 
\begin{equation}
\begin{split}
\overline{X(t_1) X(t_2) X(t_3) X(t_4)} &= \overline{X(t_1) X(t_2)} \, \overline{ X(t_3) X(t_4)} \\
&\qquad {}+ \overline{X(t_1) X(t_3)} \, \overline{ X(t_2) X(t_4)} + \overline{X(t_1) X(t_4)} \, \overline{ X(t_2) X(t_3)} \,.
\end{split}
\end{equation}
Noticing next that the second and the third terms on the right-hand-side of the latter expression provide identical contributions 
to $\overline{S(\omega_1,\TT) S(\omega_2,\TT)} $, we rewrite eq.~\eqref{SS} in the form
\begin{equation}
\begin{split}
\label{S}
\overline{S(\omega_1,\TT) S(\omega_2,\TT)} &= \frac{1}{\TT^2} \int_0^\TT \D t_1 \int^\TT_0 \D t_2 \;  \overline{X(t_1) X(t_2)} \cos\left(\omega_1 (t_1 - t_2)\right) \\
&\qquad {}\times  \int^\TT_0 \D t_3 \int^\TT_0 \D t_4 \; \overline{X(t_3) X(t_4)} \cos\left(\omega_2(t_3 - t_4)\right) \\
&\qquad {}+  \frac{2}{\TT^2} \int_0^\TT \D t_1 \int^\TT_0 \D t_2  \int^\TT_0 \D t_3 \int^\TT_0 \D t_4 \; \overline{X(t_1) X(t_3)}  \,\,\, \overline{X(t_2) X(t_4)}  \nonumber\\
&\qquad {}\times  \cos\left(\omega_1 (t_1 - t_2)\right)   \cos\left(\omega_2(t_3 - t_4)\right)  \\
&= \mu(\omega_1,\TT) \, \mu(\omega_2,\TT) + J(\omega_1,\omega_2) \,,
\end{split}
\end{equation}
where
\begin{equation}
\label{defs1}
\mu(\omega,\TT)=\overline{S(\omega,\TT)},
\end{equation}
is the mean spectral density at frequency $\omega$, while $J(\omega_1,\omega_2) $ is given by 
\begin{equation}
\begin{split}
\label{J}
J(\omega_1,\omega_2) &=  \frac{2}{\TT^2} \int_0^\TT \D t_1 \int^\TT_0 \D t_2  \int^\TT_0 \D t_3 \int^\TT_0 \D t_4 \; \overline{X(t_1) X(t_3)}  \,\,\, \overline{X(t_2) X(t_4)} \\
&\qquad{}\times  \cos\left(\omega_1 (t_1 - t_2)\right)   \cos\left(\omega_2(t_3 - t_4)\right)  \\
&= 2 W_{\cc}^2\left(\omega_1,\omega_2;\TT\right) + 2 W_{\ss}^2\left(\omega_1,\omega_2;\TT\right)\\
&\qquad {} + 2 W_{\cs}^2\left(\omega_1,\omega_2;\TT\right) + 2 W_{\sc}^2\left(\omega_1,\omega_2;\TT\right) \,.
\end{split}
\end{equation}
Here we have defined
\begin{equation}
\begin{split}
\label{W}
W_{\cc}\left(\omega_1,\omega_2;\TT\right) &= \frac{1}{\TT} \int^\TT_0 \D t_1  \int^\TT_0 \D t_2 \; \overline{X(t_1) X(t_2)} \cos\left(\omega_1 t_1\right) \cos\left(\omega_2 t_2\right) \,,  \\
W_{\ss}\left(\omega_1,\omega_2;\TT\right) &=  \frac{1}{\TT} \int^\TT_0 \D t_1  \int^\TT_0 \D t_2 \; \overline{X(t_1) X(t_2)} \sin\left(\omega_1 t_1\right) \sin\left(\omega_2 t_2\right) \,,  \\
W_{\cs}\left(\omega_1,\omega_2;\TT\right) &= \frac{1}{\TT} \int^\TT_0 \D t_1  \int^\TT_0 \D t_2 \; \overline{X(t_1) X(t_2)} \cos\left(\omega_1 t_1\right) \sin\left(\omega_2 t_2\right) \,, \\
W_{\sc}\left(\omega_1,\omega_2;\TT\right) & = \frac{1}{\TT} \int^\TT_0 \D t_1  \int^\TT_0 \D t_2 \; \overline{X(t_1) X(t_2)} \sin\left(\omega_1 t_1\right) \cos\left(\omega_2 t_2\right) \,.
\end{split}
\end{equation}
We stress that the expressions \eqref{S} to \eqref{W} hold for arbitrary Gaussian processes. Correspondingly, the Pearson correlation coefficient of the random variables $S(\omega_1,\TT)$ and $S(\omega_2,\TT)$, defined by 
\begin{align}
\label{pearson}
\rho(\omega_1,\omega_2) = \frac{\overline{S\left(\omega_1,\TT\right) S\left(\omega_2,\TT\right)}  - \mu(\omega_1, \TT) \mu(\omega_2, \TT)}{\sqrt{\left(\overline{S^2\left(\omega_1,\TT\right)} -  \mu^2(\omega_1, \TT) \right) \left(\overline{S^2\left(\omega_2,\TT\right)} -  \mu^2(\omega_2, \TT) \right)}} \,,
\end{align}
is given by
\begin{align}
\label{rhor}
\rho(\omega_1,\omega_2)= \frac{J(\omega_1,\omega_2)}{\sqrt{J(\omega_1,\omega_1) J(\omega_2,\omega_2)}} \,,
\end{align}
an expression which also holds for an arbitrary Gaussian process. 

Lastly, we note that the covariance function in the frequency domain of the random amplitude $b_\omega$ defined in 
 \eqref{prob3}  can be formally expressed through the Pearson correlation coefficient via the relation 
 \begin{align}
 \label{bb}
\overline{b_{\omega_1} b_{\omega_2}} =  \frac{\overline{S(\omega_1,\TT) S(\omega_2,\TT)}}{\mu(\omega_1,\TT) \mu(\omega_2,\TT)} =
1 + \gamma(\omega_1,\TT) \gamma(\omega_2,\TT) \rho(\omega_1,\omega_2) \,,
 \end{align} 
 where $\gamma(\omega,\TT)$ is the coefficient of variation defined in eq.~\eqref{gamma}. As we show in what follows, $\overline{b_{\omega_1} b_{\omega_2}}$ attains distinctly different values for Gaussian sub-diffusive processes, Brownian motion and super-diffusive processes, and thus offers an interesting criterion that permits to distinguish between these three cases from the analysis of correlations of $b$ in the frequency domain.

\section{Brownian motion and the Ornstein-Uhlenbeck process}
\label{BM}

As a warming-up exercise, we start with two ``simple'' cases, namely,  the 
standard Brownian motion (BM) and the Ornstein-Uhlenbeck (OU) process, for which
we can obtain comparatively compact expressions that hold for arbitrary frequencies and observation times.

\subsection{Brownian motion}
The trajectory $X(t)$ of a BM satisfies the stochastic Langevin equation
\begin{align}
\label{BMeq}
\dot{X} &=  \zeta(t) \, ,
\end{align}
where the dot stands for the time derivative, the viscosity is set for simplicity equal to unity,  
 and $\zeta$ is a Gaussian zero-mean white noise satisfying
\begin{align}
\label{noise}
\overline{\zeta(t)} = 0, \quad \overline{\zeta(t) \zeta(t')} = 2 k_\mathrm{B} T\, \delta(t-t') \,, 
\end{align} 
where $k_\mathrm{B}$ is the Boltzmann constant (set to 1 from now on), 
 and $\delta(t)$ is Dirac's delta function. Equations~\eqref{BMeq} and~\eqref{noise} imply that 
  the two-time correlation function $\overline{X(t_1) X(t_2)}$ obeys 
\begin{align}
\label{BMcor}
\overline{X(t_1) X(t_2)} = 2 \, T \, \min(t_1,t_2) \,.
\end{align}
As a consequence, for the BM, the standard spectral density $\mu(\omega,\TT)$, defined in eq.~\eqref{defs1},  is given for arbitrary $\omega$ and $\TT$  by 
\begin{align}
\mu(\omega,\TT) = \frac{4 T}{\omega^2} \left[1 - \frac{\sin(\omega \TT)}{\omega \TT}\right] \, \qquad \mu(\omega) =\lim_{\TT\to\infty}\mu(\omega,\TT)= \frac{4 T}{\omega^2} \,.
\end{align}
See, e.g. \cite{krapf1} for more details.

\begin{figure}
\begin{center}
\includegraphics[width=150mm]{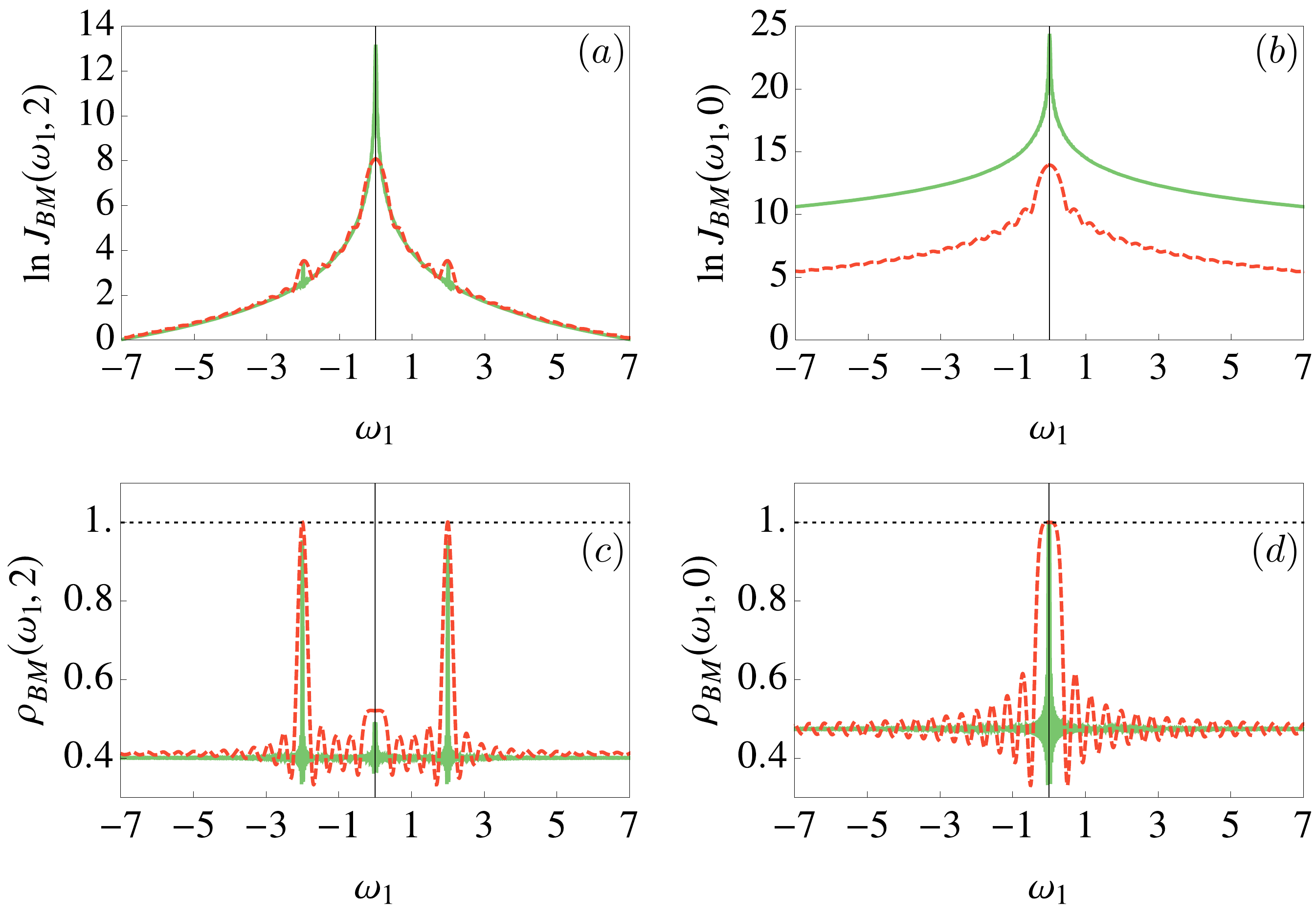}
\end{center}
\caption{Brownian motion. Logarithm of $J_{\mathrm{BM}}(\omega_1,\omega_2)$ (upper panels) in eq.~\eqref{JBM} and  Pearson correlation coefficient $\rho_{\mathrm{BM}}(\omega_1,\omega_2)$ (lower panels)
as functions of $\omega_1$ for fixed $\omega_2=2$ (left panels) and $\omega_2 =0 $ (right panels). Temperature: $T = 5$. Solid lines: $\TT = 200$. Dashed lines: $\TT =15$.
}
\label{fig1}
\end{figure}

Next, inserting the expression \eqref{BMcor} into eq.~\eqref{J} and integrating, we find the following exact expression, which is valid for arbitrary frequencies and observation times: 
\begin{equation}
\begin{split}
\label{JBM}
J_{\mathrm{BM}}(\omega_1,\omega_2) &= \frac{8 T^2}{\omega_1^2 \omega_2^2} \left[1  - \frac{2 \sin\left(\omega_1 \TT\right)}{\omega_1 \TT} - \frac{2 \sin\left(\omega_2 \TT\right)}{\omega_2 \TT} \right.\\
&\qquad {}+ \frac{2 \left(\omega_1 \sin\left(\omega_1 \TT\right)  \cos\left(\omega_2 \TT\right) - \omega_2 \sin\left(\omega_2 \TT\right)  \cos\left(\omega_1 \TT\right)\right)}{\left(\omega_1^2 - \omega_2^2\right) \TT}  \\ 
&\qquad {}+ \frac{2 \left(\omega_1^6+\omega_2^6\right)}{\omega_1^2 \omega_2^2 \left(\omega_1^2 - \omega_2^2\right)^2 \TT^2} - 
 \frac{2 \left(\omega_1^2 + \omega_2^2\right) \left(\omega_1^2 \cos\left(\omega_2 \TT\right) - \omega_2^2 \cos\left(\omega_1 \TT\right)\right)}{\omega_1^2 \omega_2^2 \left(\omega_1^2 - \omega_2^2\right) \TT^2} \\
 &\qquad \left. {}- \frac{2 \left(\left(\omega_1^2 + \omega_2^2\right)  \cos\left(\omega_1 \TT\right) \cos\left(\omega_2 \TT\right) + 2 \omega_1 \omega_2 \sin\left(\omega_1 \TT\right) \sin\left(\omega_2 \TT\right)\right)}{\left(\omega_1^2 - \omega_2^2\right)^2 \TT^2}
\right] \,.
\end{split}
\end{equation}
We show in~fig.~\ref{fig1} (upper panels) $J_{\mathrm{BM}}(\omega_1,\omega_2)$ as a function of $\omega_1$ for  $\omega_2 =  2$ and $\omega_2 =0$, and for $\TT = 15$ and $\TT = 200$. The $y$-axis is logarithmic since the variation range of $J_{\mathrm{BM}}$ is large, in particular as a function of $\TT$ when one of the frequencies vanishes. 

The behaviour of $J_{\mathrm{BM}}(\omega_1,\omega_2)$ is rather rich and deserves a detailed discussion. First of all, $J_{\mathrm{BM}}(\omega_1,\omega_2)$ is an oscillatory and symmetrical function of both frequencies. When, e.g., $\omega_2$ is kept fixed at a nonzero value,  $J_{\mathrm{BM}}(\omega_1,\omega_2)$ exhibits pronounced peaks at $\omega_1=0$ and  at $\omega_1 = \pm \omega_2$ (see panels (a) and (b) in Fig.~\ref{fig1}). In this case, the height 
of the two peaks at $\omega_1 =\pm  \omega_2 = \omega$ is given by 
\begin{equation}
\begin{split}
\label{BMff}
J_{\mathrm{BM}}(\omega,\omega) &= \frac{20 T^2}{\omega^4} \Bigg[1 + \frac{\sin\left(2 \omega \TT\right)}{5 \omega \TT} - \frac{12 \sin\left(\omega \TT\right)}{5 \omega \TT}\\
&\qquad\qquad  {} +  \frac{17}{10 \omega^2 \TT^2} -  \frac{8 \cos\left(\omega \TT\right)}{5 \omega^2 \TT^2}  -  \frac{\cos\left(2 \omega \TT\right)}{10 \omega^2 \TT^2}
\Bigg] \,.
\end{split}
\end{equation}
This expression approaches $20\, T^2/\omega^4$ as $\TT \to \infty$. Consequently, the height of the peak at $\omega_1=\pm \omega_2$ vanishes as $\omega^{-4}$ when $\omega \to \infty$. In turn, the height of the peak at $\omega_1 = 0$ (for $\omega_2 \neq 0$) is given by
\begin{equation}
\begin{split}
\label{22}
J_{\mathrm{BM}}(0,\omega_2) &= \frac{2 T^2 \, \TT^2}{\omega_2^2}  \Bigg[1 - \frac{4 \sin\left(\omega_2 \TT\right)}{\omega_2 \TT} + \frac{8}{\omega_2^2 \TT^2} \\
&\qquad {}- \frac{4 \cos\left(\omega_2 \TT\right)}{\omega^2 \TT^2} - \frac{8 \sin\left(\omega_2 \TT\right)}{\omega_2^3 \TT^3} + \frac{8}{\omega_2^4 \TT^4} - \frac{8 \cos\left(\omega_2 \TT\right)}{\omega^4 \TT^4}
 \Bigg] \,.
\end{split}
\end{equation}
Consequently, in the leading order in~$\TT$, $J_{\mathrm{BM}}(0,\omega_2)$ 
grows proportionally to~$\TT^2$ and decreases as the frequency $\omega_2$ grows. On the other hand, when $\omega_2 = 0$ there is a single peak at $\omega_1 = 0$ (see panel (b) in Fig.~\ref{fig1}), whose height  is given by
\begin{align}
\label{23}
J_{\mathrm{BM}}(0,0) = \frac{8 T^2}{9}  \TT^4 \,,
\end{align}
i.e., it is a monotonically increasing function of the observation time $\TT$.

We now discuss the leading large-$\TT$ behaviour of the Pearson correlation coefficient $\rho_{\mathrm{BM}}(\omega_1,\omega_2)$ (see panels (c) and (d) in Fig.~\ref{fig1}). For $\omega_2 \neq 0$, we obviously have $\rho_{\mathrm{BM}}(\omega_1,\omega_2) = 1$
at $\omega_1 = \pm \omega_2$. On the other hand,  
$\rho_{\mathrm{BM}}(\omega_1,\omega_2)$ does not vanish for $\omega_1 \neq \omega_2$ when neither frequency vanishes, but approaches in the large-$\TT$ limit the constant value
\begin{align}
\rho_{\mathrm{BM}}(\omega_1,\omega_2) = \frac{2}{5} \,.
\end{align}
This means that 
the single-trajectory PSDs $S(\omega_1,\TT \to \infty)$ and $S(\omega_2,\TT \to \infty)$ remain correlated for arbitrary different values $\omega_1$ and $\omega_2$. Moreover, the Pearson coefficient exhibits an additional peak at $\omega_1\neq 0$, whose value approaches
\begin{align}
\label{funny}
\rho_{\mathrm{BM}}(\omega_1=0,\omega_2 \neq 0) = \frac{3}{2 \sqrt{10}} \approx 0.474, 
\end{align}
in the limit $\TT \to \infty$.
Consequently, there is some excess correlation between $S(\omega,\TT \to \infty)$ for non-vanishing $\omega$ and $S(0,\TT \to \infty)$. We note in passing that $S(0,\TT)$ is formally given by the squared area under the random curve $X(t)$, divided by $\TT$. 

\begin{figure}
\begin{center}
\includegraphics[width=140mm]{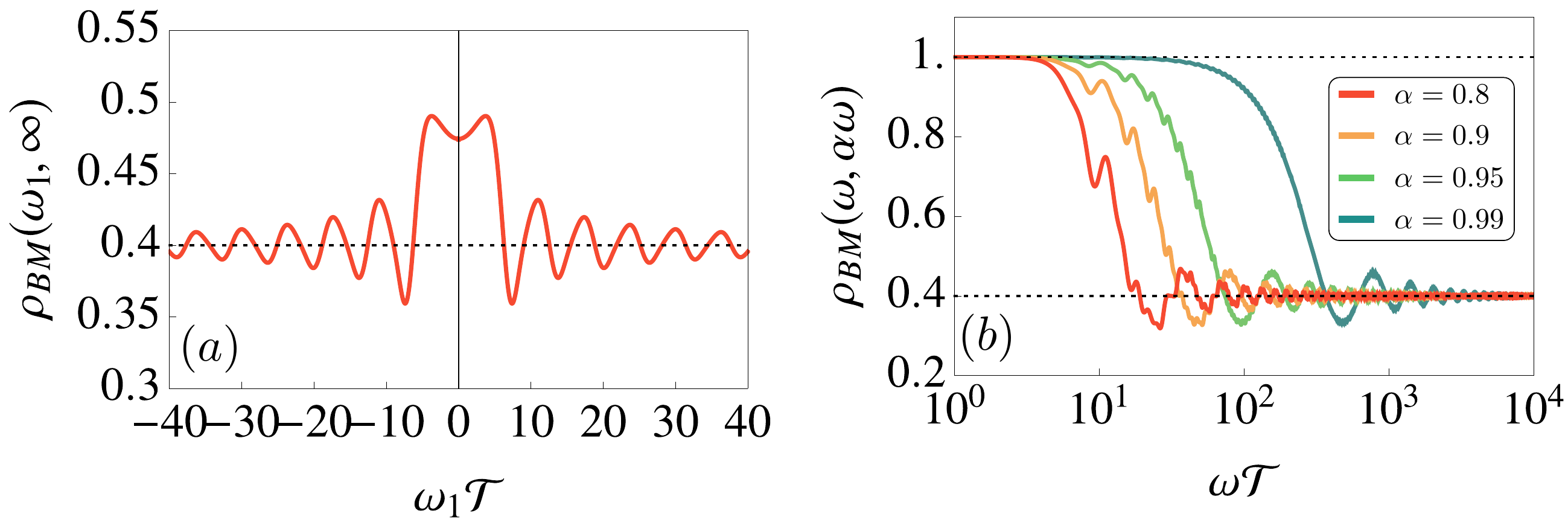}
\end{center}
\caption{Brownian motion. Panel (a): Pearson correlation coefficient $\rho_{\mathrm{BM}}(\omega_1,\omega_2\to \infty)$, eq.~\eqref{infty}, as a function of $\omega_1 \TT$. Panel (b): Pearson correlation coefficient $\rho_{\mathrm{BM}}(\omega_1,\omega_2)$ for $\omega_1 =\omega$ and $\omega_2 = \alpha \omega$ as a function of $\omega \TT$ for several values of the parameter $\alpha$. In both panels the horizontal dashed line indicates the constant level $2/5$.
}
\label{fig2}
\end{figure}

Further on, we consider the behaviour at  finite observation time and focus on two special cases:  
a)
$\omega_1$ is
 fixed, while $\omega_2 \to \infty$, (i.e., we quantify the correlations between $S(\omega_1,\TT)$ and $S(\infty, \TT)$), and b) $\omega_1 = \omega$ while $\omega_2 = \alpha \, \omega$, where $\alpha \leq 1$ is a scale parameter. 
In the case a), the Pearson correlation coefficient is given by 
\begin{equation}
\begin{split}
\label{infty}
&\rho_{\mathrm{BM}}(\omega_1,\omega_2 \to \infty) = \frac{J_{\mathrm{BM}}(\omega_1,\infty)}{\sqrt{J_{\mathrm{BM}}(\omega_1,\omega_1) J_{\mathrm{BM}}(\infty,\infty)}} \\
&\qquad = \frac{2}{5} \dfrac{1 - \dfrac{2 \sin(\omega_1 \TT)}{\omega_1 \TT} + \dfrac{2 \left(1 - \cos(\omega_1 \TT)\right)}{\omega_1^2 \TT^2}}{\sqrt{1 + \dfrac{\sin(2 \omega_1 \TT)- 12 \sin(\omega_1 \TT)}{5 \omega_1 \TT} + \dfrac{17 - 16 \cos(\omega_1 \TT) -\cos(2 \omega_1 \TT)}{10 \omega_1^2 \TT^2} }} \,,
\end{split}
\end{equation}
and hence is a function of the product $\omega_1 \TT$ only. Its behavior is shown in fig.~\ref{fig2}, panel (a) as a function of the product $\omega_1 \TT$. One can see that the behavior of the correlations is quite complicated and that in particular $S(\omega,\TT)$ and $S(\infty,\TT)$ do not decouple neither at a finite nor at an infinite observation time. 
The case b) is shown in fig.~\ref{fig2}, panel (b), where $\rho_{\mathrm{BM}}(\omega, \alpha \, \omega)$ is plotted as a function of $\omega \TT$ for several values of $\alpha$.  We observe that  $S(\omega,\TT)$ and $S(\alpha \, \omega,\TT)$ are completely correlated within some well-defined 
region of values of $\omega \TT$, and such a correlation drops rapidly to the constant level $2/5$ upon a further increase of $\omega \TT$. The size of the window in which $S(\omega,\TT)$ and $S(\alpha \omega,\TT)$ are completely correlated increases as $\alpha$ approaches $1$, which is not counter-intuitive.

Finally, summarizing our results for the BM and recalling that $\gamma_{\mathrm{BM}}(0,\TT) = \sqrt{2}$ and $\gamma_{\mathrm{BM}}(\omega>0,\TT=\infty) = \sqrt{5}/2$
\cite{krapf1}, we find that the covariance function of the random amplitude $b_\omega$ (see eqs.~\eqref{prob3}  and \eqref{bb}) in the frequency domain is given, as $\TT\to\infty$, by
\begin{equation}
\label{bbBM}
\overline{b_{\omega_1} b_{\omega_2}} =  
\begin{cases}
3/2  \, ,     & \text{for } \omega_1 \neq \omega_2, \quad \omega_1,\omega_2 > 0; \\
7/4  \, ,  &\text{for }\omega_1=0, \quad \omega_2 > 0 ;   \\
9/4\,, &\text{for }\omega_1=\omega_2 > 0;\\
3 \,, &\text{for } \omega_1 = \omega_2 = 0 .
\end{cases}
\end{equation}
This implies that the $b_\omega$ exhibit some degree of correlation for different values of~$\omega$.

\subsection{Ornstein-Uhlenbeck process}
\label{OUS}

The OU process obeys the Langevin equation
\begin{align}
\label{OUdef}
\dot{X} &= - X + \zeta(t) \,, 
\end{align}
where $\zeta(t)$ is a Gaussian noise whose properties are defined in eq.~\eqref{noise}. At variance with the BM, this process approaches a stationary distribution when $t\to\infty$. The solution of eq.~\eqref{OUdef} for the initial condition $X(0)=0$ and a given realization of the noise is given by
\begin{align}
X(t)&= \E^{-t} \int^t_0\D\tau\; \E^{\tau} \, \zeta(\tau)  \,.
\end{align}
Consequently, the two-time correlation function of $X(t)$ for $t_2 \leq t_1$ is given~by
\begin{align}
\label{OUcor}
\overline{X(t_1) X(t_2)} = T \left[\exp\left(-(t_1 - t_2)\right) - \exp\left(- (t_1 + t_2)\right) \right] \,.
\end{align}
The correlation function $\overline{X(t_1) X(t_2)} $ for $t_2 \geq t_1$ is obtained by merely interchanging $t_2$ and $t_1$ in this relation. Note that, in contrast to the BM whose mean-squared displacement grows without bounds in time, the variance of OU process is bounded. Indeed, setting $t_1=t_2 =t$ in eq.~\eqref{OUcor}, we obtain $\overline{X^2(t)} = T \left(1 - \exp(- 2 t)\right) $,  and therefore  the mean-squared displacement from the origin approaches the constant value $T$ for $t \to \infty$.

\begin{figure}
\begin{center}
\includegraphics[width=140mm]{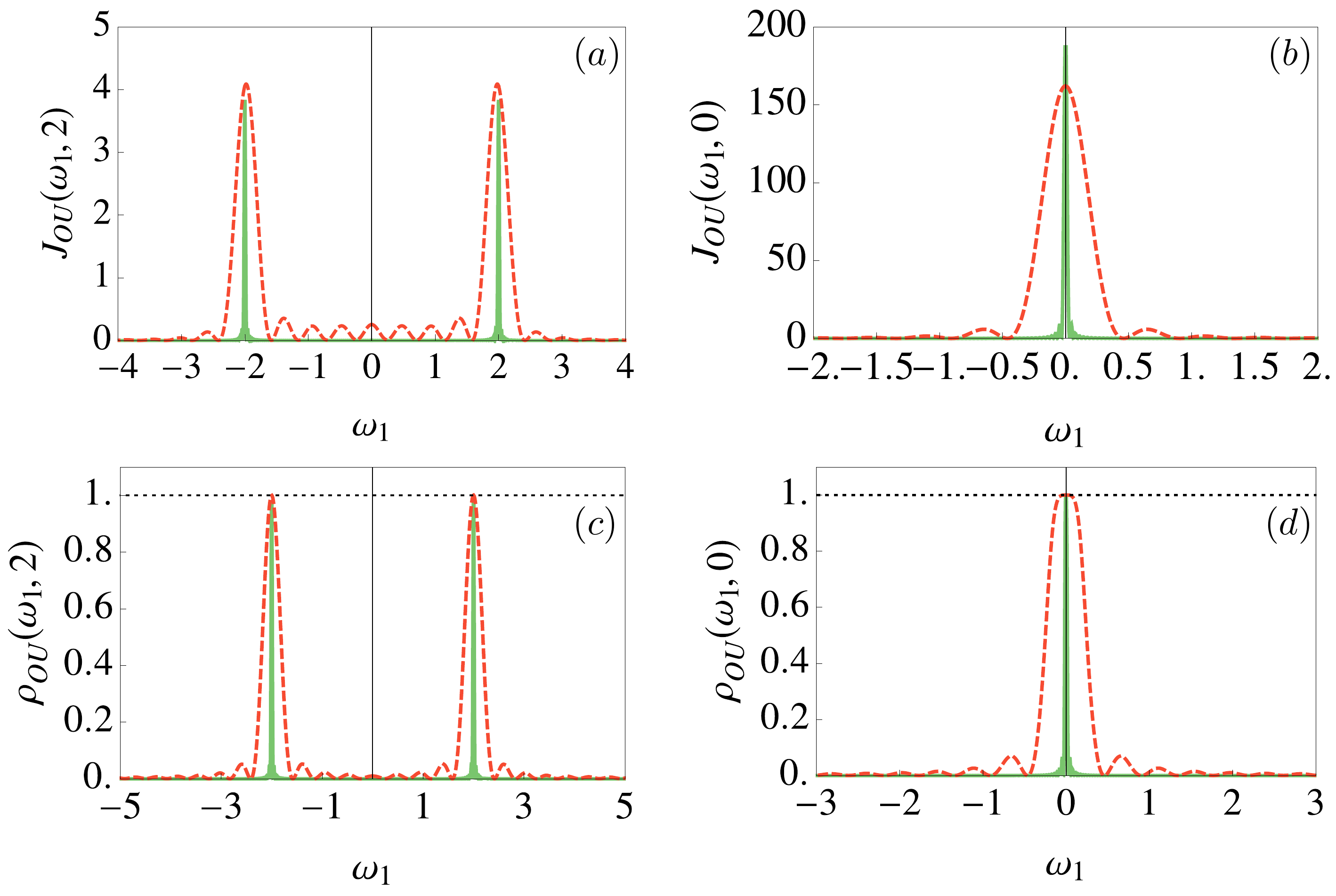}
\end{center}
\caption{Ornstein-Uhlenbeck process. The excess correlation function $J_{\mathrm{OU}}(\omega_1,\omega_2)$ and the Pearson correlation coefficient $\rho_{\mathrm{OU}}(\omega_1,\omega_2)$ are plotted
as functions of $\omega_1$ for fixed $\omega_2=2$ (panels (a) and (c)) and $\omega_2 =0 $ (panels (b) and (d)). The temperature $T$ is set to $T5$. Solid lines depict $J_{\mathrm{OU}}(\omega_1,\omega_2)$ and $\rho_{\mathrm{OU}}(\omega_1,\omega_2)$ for $\TT = 200$, and the dashed lines for $\TT =15$.
}
\label{fig3}
\end{figure}

Inserting the expression \eqref{OUcor} into eqs.~\eqref{W} and performing the integrations, we eventually find a rather lengthy expression for $J_{\mathrm{OU}}(\omega_1,\omega_2) $, which holds for arbitrary $\omega_1$, $\omega_2$, and $\TT$ and it is unnecessary to copy here. We show in fig.~\ref{fig3} the behavior of
$J_{\mathrm{OU}}(\omega_1,\omega_2)$ as a function of $\omega_1$ for two values of the frequency $\omega_2$: $\omega_2=2$ and $\omega_2=0$ (see panels (a) and (b)).  We observe that $J_{\mathrm{OU}}(\omega_1,\omega_2) $  behaves differently from the BM case in several aspects. First of all, for $\omega_2 = 2$ the excess frequency-frequency correlation function exhibits only the peaks at $\omega_1 = \pm \omega_2$, while the peak at $\omega_1=0$ is absent. Moreover, the height of the peak at $\omega_1 = \pm \omega_2 \neq 0$ is given by
\begin{equation}
\begin{split}
\label{zz}
J_{\mathrm{OU}}(\omega,\omega) &= \frac{4 T^2}{(1+\omega^2)^2} + \frac{4  T^2 \left(\omega^2 - 3\right)}{\left(1 + \omega^2\right)^3 \TT} + \frac{2  T^2 \left(1 + 7 \omega^2 - \omega^4 + \omega^6\right)}{\omega^2 (1+\omega^2)^4 \TT^2} \\
&\qquad {}+ \frac{2  T^2 \left(3 \omega^2 - 1\right) \cos\left(2 \omega \TT\right)}{\omega^2 (1+\omega^2)^3 \TT^2} + \frac{2 T^2 \left(\omega^2 - 3\right) \sin\left(2 \omega \TT\right)}{\omega (1+\omega^2)^3 \TT^2}  + \mathcal{O}\left(\E^{-\TT}\right) \,,
\end{split}
\end{equation}
where the symbol $\mathcal{O}\left(\E^{-\TT}\right)$ signifies that the omitted terms decay exponentially with the observation time. Therefore, the height of the peaks approaches a $\TT$-independent value as $\TT \to \infty$, which decays as the fourth inverse power of the frequency in the limit $\omega \to \infty$. This is quite similar to the behaviour of $J_{\mathrm{BM}}(\omega,\omega)$ (see eq.~\eqref{BMff}). Note that the limit $\omega \to 0$ of the expression \eqref{zz} does not correctly reproduce $J_{\mathrm{OU}}(0,0)$. In fact, 
the height of the peak in case $\omega_1=\omega_2=0$ must be evaluated by setting $\omega_1=0$ and $\omega_2=0$ from the very beginning. In doing so, we find that $J_{\mathrm{OU}}(0,0)$ is given by
\begin{align}
\label{z3}
J_{\mathrm{OU}}(0,0) = 8 T^2 - \frac{24 T^2}{\TT} + \frac{18 T^2}{\TT^2} + \mathcal{O}\left(\E^{-\TT}\right) \,,
\end{align}
where the coefficient in front of $T^2$ is twice what would be obtained directly from  eq.~\eqref{zz}.

We now dwell on the case $\omega_1 \neq \omega_2$. Here, rather lengthy calculations show that $J_{\mathrm{OU}}(\omega_1,\omega_2)$ has the following large-$\TT$ asymptotic form:
\begin{align}
\label{OU}
&J_{\mathrm{OU}}(\omega_1,\omega_2) = \frac{2 T^2 A_{\mathrm{OU}}(\omega_1,\omega_2)}{\TT^2} + O\left(\E^{-\TT}\right) \,,
\end{align}
where the omitted terms decay, in the leading order, in proportion to $\exp(-\TT)$, while the decay amplitude $A_{\mathrm{OU}}(\omega_1,\omega_2)$ in the first term on the right-hand-side of eq.~\eqref{OU} is given explicitly by
\begin{equation}
\begin{split}
\label{AOU}
A_{\mathrm{OU}}(\omega_1,\omega_2) &= \frac{8 \left(\omega_1^2 +  \omega_2^2\right) +\left(\omega_1^2 - \omega_2^2\right)^2}{\left(1 + \omega_1^2\right) \left(1 + \omega_2^2\right)\left(\omega_1^2 - \omega_2^2\right)^2} \\
&\qquad {} - \frac{4 \left(\left(1 + \omega_1 \omega_2\right)^2 + \omega_1 \omega_2 \left(\omega_1 - \omega_2\right)^2\right)}{\left(1 + \omega_1^2\right)^2 \left(1 + \omega_2^2\right)^2\left(\omega_1 - \omega_2\right)^2} \cos\left(\left(\omega_1 - \omega_2\right) \TT\right) \\
&\qquad {}- 
\frac{4 \left(\left(1 - \omega_1 \omega_2\right)^2 - \omega_1 \omega_2 \left(\omega_1 + \omega_2\right)^2\right)}{\left(1 + \omega_1^2\right)^2 \left(1 + \omega_2^2\right)^2\left(\omega_1 + \omega_2\right)^2} \cos\left(\left(\omega_1 + \omega_2\right) \TT\right) \\
&\qquad {}- \frac{2 \left(3 + \omega_1^2 + \omega_2^2 - \omega_1^2 \omega_2^2\right)}{\left(1+\omega_1^2\right)^2 \left(1 + \omega_2^2\right)^2} \frac{\sin\left((\omega_1 - \omega_2) \TT\right)}{\omega_1-\omega_2} \\
& \qquad {}- \frac{2 \left(3 + \omega_1^2 + \omega_2^2 - \omega_1^2 \omega_2^2\right)}{\left(1+\omega_1^2\right)^2 \left(1 + \omega_2^2\right)^2} \frac{\sin\left((\omega_1 + \omega_2) \TT\right)}{\omega_1 + \omega_2} \,.
\end{split}
\end{equation}
This expression contains the $\TT$-independent first term, while the following ones oscillate with $\TT$. 
Therefore, when $\omega_1\neq \omega_2$ and both are non-vanishing, $J_{\mathrm{OU}}(\omega_1,\omega_2)$ vanishes in the limit $\TT \to \infty$, implying  that the random variables $S(\omega_1,\TT \to \infty)$ and   $S(\omega_2,\TT \to \infty)$  become statistically independent.  Moreover, keeping $\omega_1$ fixed  and $\omega_2 \to \infty$,  we observe that $A_{\mathrm{OU}}(\omega_1,\omega_2)$ in eq.~\eqref{AOU} vanishes as $1/\omega_2^2$, meaning that correlations between $S(\omega_1,\TT)$ and $S(\omega_2,\TT)$ also vanish when $\TT$ is finite (but large enough to ensure the validity of the asymptotic form in eq.~\eqref{OU}) and  $\abs{\omega_1 - \omega_2}\to \infty$. Such a behaviour is markedly different from the one we found for the BM. Finally, we obtain that the covariance function of the random amplitude $b_\omega$ (see eq.~\eqref{prob3}) satisfies in the limit $\TT \to \infty$ the relations
\begin{equation}
\label{bbOU}
\overline{b_{\omega_1} b_{\omega_2}} =  
\begin{cases}
1  \, ,     & \text{for }\omega_1 \neq \omega_2,\quad \, \omega_1,\omega_2 > 0 ;\\
1  \, ,  &\text{for }\omega_1=0, \quad \omega_2 > 0 ;   \\
2  \, , &\text{for } \omega_1=\omega_2 >0;\\
3 \,, & \text{for }\omega_1 = \omega_2 = 0 .
\end{cases}
\end{equation}
These values are different from the BM case, except for the case $\omega_1 = \omega_2=0$. Notice that this implies that the $b_\omega$ for different values of $\omega$ are totally uncorrelated.

\section{Brownian gyrator}
\label{BG}

We focus next on the Brownian gyrator (BG) model
 \cite{pel}, defined as
a pair of coupled OU processes $X(t)$ and $Y(t)$, which obey 
the following system of  Langevin equations: 
\begin{equation}
\begin{split}
\label{G}
\dot{X} &= - X + u \, Y + \zeta_x(t) \,, \nonumber\\
\dot{Y} &= - Y + u \, X + \zeta_y(t) \,. 
\end{split}
\end{equation}
Here $\zeta_{x,y}$ are Gaussian zero-mean white noises such that  
\begin{align}
\overline{\zeta_i(t) \zeta_j(t')} = 2 T_i \delta_{i,j} \delta(t-t') \,, \quad i,j = x,y \,,
\end{align} 
where $\delta_{i,j}$ is the Kronecker delta, while $T_x$ and $T_y$ are the temperatures (measured in units of the Boltzmann constant) of two thermal baths.

Despite its simplicity, the model defined by eqs.~\eqref{G} exhibits a rather non-trivial physical behaviour.  It was in fact realized in \cite{1} that it represents a minimal model of a heat machine. When $u=0$, $X(t)$ and $Y(t)$ decouple and form two independent OU processes, leading us back to the case studied in the previous section~\ref{OUS}. On the other hand, when $0< |u| < 1$ and $T_x \neq T_y$, the system eventually reaches a non-equilibrium steady state, characterized by a non-vanishing average rotation frequency in the $(X,Y)$ plane. This explains the name of the model. 
 Various aspects of the BG dynamical behaviour and of its steady-state properties 
have been studied, see, e.g., \cite{alberto1,alberto2} and \cite{crisanti,2,12,11,3,13,Lam3,Lam4,cherail,CilMD,ital}. 
In particular,  a non-trivial fluctuation theorem was established by considering the response of the BG to external regular forces, what allowed to define 
explicitly an effective temperatures \cite{Lam3,Lam4,cherail}. We also remark that the 
setting with constant forces exerted
on the BG is mathematically identical to 
the one-dimensional bead-spring model studied via Brownian-dynamics simulations in 
 \cite{fakhri1} and analytically in \cite{fakhri2} and \cite{komura}. A generalization of the BG model 
 for a system of two coupled noisy Kuramoto oscillators has been discussed in  \cite{njp}.

 Solving 
 equations \eqref{G} for the trajectories, with the initial condition $X(0)=Y(0)=0$, we find that  for given realisations of noises $ \zeta_x(\tau)$ and  $ \zeta_y(\tau)$, $X(t)$ and $Y(t)$ are given by
 \begin{equation}
 \begin{split}
X(t)&= \E^{-t} \int^t_0 \D\tau\; \E^{\tau} \cosh\left(u (t - \tau)\right) \zeta_x(\tau) + \E^{-t} \int^t_0 \D\tau\;\E^{\tau} \sinh\left(u (t - \tau)\right) \zeta_y(\tau) \,, \\
Y(t)&= \E^{-t} \int^t_0\D\tau\; \E^{\tau} \sinh\left(u (t - \tau)\right) \zeta_x(\tau)  + \E^{-t} \int^t_0 \D\tau\;\E^{\tau} \cosh\left(u (t - \tau)\right) \zeta_y(\tau)  \,.
\end{split}
\end{equation}

We thus obtain the following expressions for the two-time correlation function of the $X$-component. For $t_1 \geq t_2$ we have
 \begin{equation}
 \begin{split}
\label{a}
&\overline{X(t_1) X(t_2)} = \frac{\E^{-t_1-t_2} }{2 (1-u^2)}  \Bigg[ \cosh\left(u(t_1-t_2)\right)\\
&\qquad {}\times
\Big(T_y + \E^{t_2} \Big(T_x \cosh(t_2) +\left(3 T_x + 2 u^2 \left(T_y - T_x\right)\right) \sinh(t_2)\Big)\Big)\\
&\qquad {} - \left(T_x+T_y\right)\\
&\qquad {}\times \Big(\cosh\left(u(t_1-t_2) \right) + u \Big(\sinh\left(u(t_1 - t_2)\right) + \E^{-2 t_2} \sinh\left(u(t_1 -t_2)\right)\Big)
\Big)
\Bigg].
\end{split}
\end{equation}
For $t_2 \geq t_1$ we have
\begin{equation}
 \begin{split}
\label{b}
&\overline{X(t_1) X(t_2)} = \frac{\E^{-t_1-t_2}}{2 (1-u^2)}   \Bigg[\cosh\left(u(t_2-t_1)\right) \\
&\qquad {}\times\Big(T_y + \E^{t_1} \Big(T_x \cosh(t_1) + \left(3 T_x + 2 u^2 \left(T_y - T_x\right)\right) \sinh(t_1)\Big)\Big)\\
&\qquad {} -\left(T_x + T_y\right) \\
&\qquad {}\times\Big(\cosh\left(u (t_1 + t_2)\right) +   u \Big(\sinh\left(u (t_1 + t_2)\right) - \E^{2 t_1} \sinh\left(u(t_2 - t_1)\right) \Big)\Big)
\Bigg] \,.
\end{split}
\end{equation}
As one may readily check, the variance $\overline{X^2(t)}$ of the $X$-component is bounded in the $t \to \infty$ limit:
\begin{align}
\overline{X^2(t)} = \frac{T_x + (u^2/2) (T_y-T_x)}{1-u^2} + \mathcal{O}\left(\E^{-(1 - |u|) t}\right) \,.
\end{align} 
The latter expression naturally reduces to the one obtained for the OU process when $u = 0$. A similar expression holds for the $Y$ component.

We now consider the PSD of the $X$-component of the BG. First, by
inserting the expressions \eqref{a} and \eqref{b} into eqs.~\eqref{defs1} and \eqref{def2}, and evaluating the integrals, we obtain the following large-$\TT$ form for
the mean PSD of the $X$-component:
\begin{equation}
\begin{split}
\label{mean}
&\mu_x(\omega,\TT) \simeq \frac{2 \Big(\left(1+\omega^2\right) T_x + u^2 T_y\Big)}{\left(\omega^2 + (1 - u)^2\right) \left(\omega^2 + (1+u)^2\right)} \\
&\qquad {}- \frac{2 \Big(\left(\omega^4 + \left(1+\omega^2\right)^2 - (u^2+\omega^2)^2\right) T_x + 2 u^2 (1-u^2) T_y + 2 u^2 \omega^2 \left(T_y - T_x\right)\Big)}{\left(\omega^2 + (1 - u)^2\right)^2 \left(\omega^2 + (1+u)^2\right)^2} \frac{1}{\TT} \,,
\end{split}
\end{equation}
where the omitted terms decay exponentially with $\TT$ in the limit $\TT \to \infty$. The symbol $\simeq$ here and henceforth signifies that we consider only the leading terms in the limit $\TT \to \infty$.
\begin{figure}
\begin{center}
\includegraphics[width=140mm]{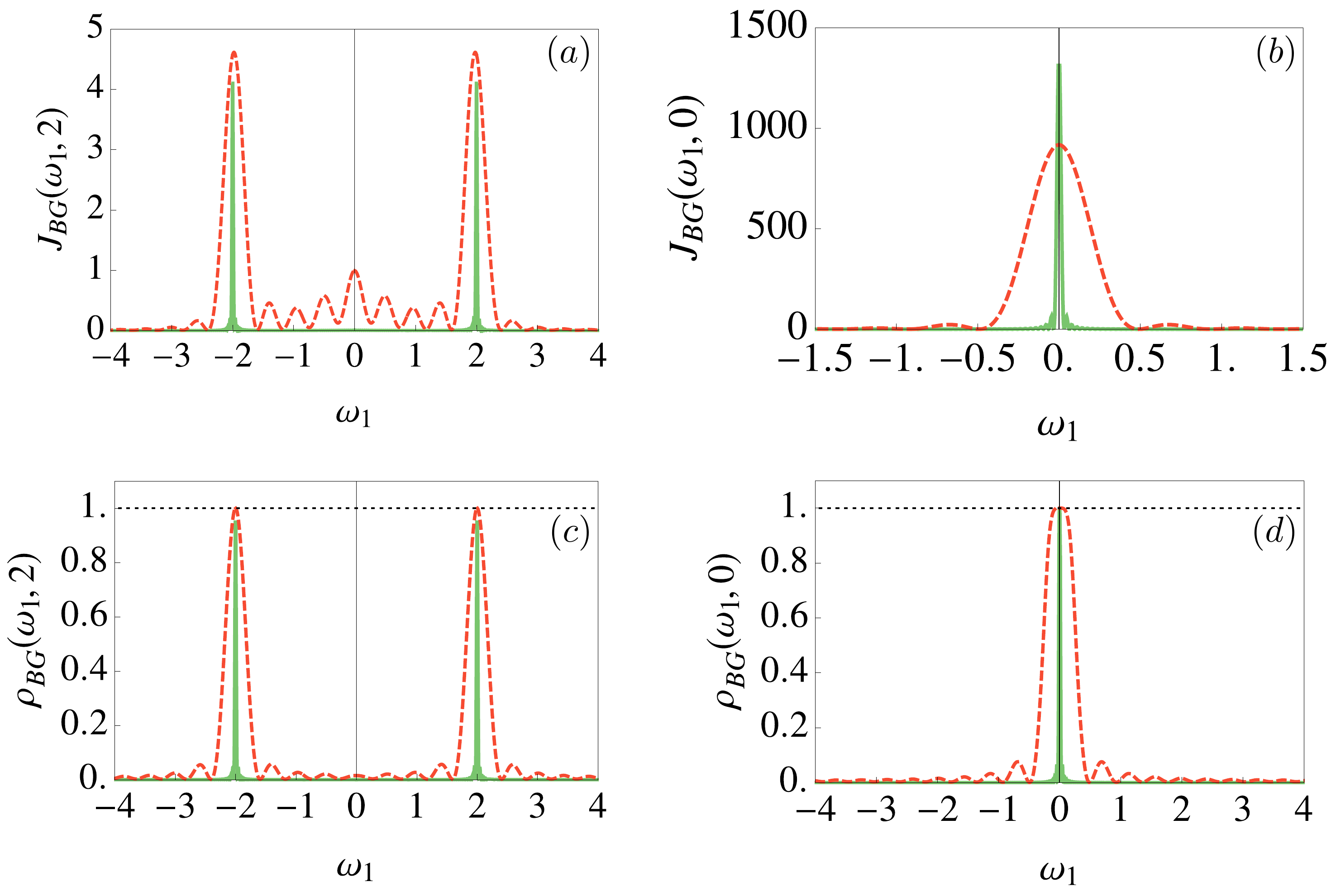}
\end{center}
\caption{Brownian gyrator with the temperatures $T_x = 5$ and $T_y = 10$. The excess correlation function  $J_{\mathrm{BG}}(\omega_1,\omega_2)$ (upper panels) and the Pearson correlation coefficient $\rho_{\mathrm{BG}}(\omega_1,\omega_2)$ (lower panels) as functions of $\omega_1$ for fixed $\omega_2=2$ (left panels) and $\omega_2 =0 $ (right panels). Solid lines: $\TT = 200$; dashed lines: $\TT =15$.
}
\label{fig40}
\end{figure}
Next, inserting \eqref{a} and \eqref{b} into eq.~\eqref{J}, we evaluate the excess correlation 
function $J_{\mathrm{BG}}(\omega_1,\omega_2)$ for arbitrary $\TT$, $\omega_1$, and $\omega_2$. Since the obtained expression appears to be too lengthy to be listed explicitly here, we instead show its behavior it in Fig.~\ref{fig40} in the same way as we did it for the OU process. In panels (a) and (b) we present 
$J_{\mathrm{BG}}(\omega_1,\omega_2)$ as function of $\omega_1$ for $\TT =15$ and $\TT =200$, and two fixed values of $\omega_2$: $\omega_2 = 2$ and $\omega_2 =0$. We observe that $J_{\mathrm{BG}}(\omega_1,\omega_2)$ exhibits essentially the same behaviour as $J_{\mathrm{OU}}(\omega_1,\omega_2)$, which was discussed 
in the previous section, and differs from it only in some details. The excess correlation function $J_{\mathrm{BG}}(\omega_1,\omega_2)$
is an oscillatory function  of $\omega_1$ (and hence, of $\omega_2$) 
with peaks at $\omega_1 = \pm \omega_2$, which naturally merge into a single peak for $\omega_2 =0$ (see panel (b)). The height of the
peaks, i.e., $J_{\mathrm{BG}}(\omega_1=\omega_2=\omega)$ (for $\omega$ bounded away from zero) is given in the limit $\TT \to \infty$
by
\begin{align}
J_{\mathrm{BG}}(\omega_1=\omega_2=\omega) = \frac{\left( 2 (1-u^2)^2 \left(4 (1+\omega^2) T_x + 3 u^2 T_y\right)^2 + u^4 \left(\omega^6+2(1+u^2)^2\right) T_y^2    \right)}{8 \left(1-u^2\right)^2 \left(\omega^4 + 2 (1+u^2) \omega^2 + (1-u^2)^2\right)^2} \,,
\end{align}
i.e., is $\TT$-independent as is its counterpart for the OU process. The difference with the OU case, (apart from 
the fact that $J_{\mathrm{BG}}(\omega,\omega)$ depends on the parameters $u$, $T_x$ and $T_y$), is that the height of the peaks vanishes with $\omega$ much slower, as the second inverse power of the frequency, 
\begin{align}
J_{\mathrm{BG}}(\omega_1=\omega_2=\omega) = \frac{u^4 T_y^2}{8 (1-u^2)^2 \omega^2} + O\left(\frac{1}{\omega^4}\right) \,,
\end{align}
while $J_{\mathrm{OU}}(\omega_1=\omega_2=\omega) \sim 1/\omega^4$ (see eq.~\eqref{zz}). 
Interestingly enough, the amplitude in this large-$\omega$ asymptotic form for the $X$-component is entirely defined by the temperature of the $Y$-component of the BG and the coupling parameter $u$. In turn, for $\omega_2 = 0$, the height of the peak at $\omega_1=0$ (see panel (b) in Fig. \ref{fig40}) approaches in the asymptotic limit $\TT \to \infty$ the following constant value
\begin{align}
J_{\mathrm{BG}}(\omega_1=\omega_2=0) = \frac{8 \left(T_x + u^2 T_y\right)^2}{(1-u^2)^4} \,,
\end{align} 
which reduces to the one in eq.~\eqref{z3} when $u=0$, i.e., the  $X$- and $Y$-components of the BG decouple.

Lastly, 
analysing the obtained expression for $J_{\mathrm{BG}}(\omega_1,\omega_2)$, we realize that 
for $\omega_1 \neq \omega_2$ (and both frequencies bounded away from zero) the leading behavior in the large-$\TT$ limit is given by
\begin{align}
J_{\mathrm{BG}}(\omega_1,\omega_2) = \frac{A_{\mathrm{BG}}(\omega_1,\omega_2)}{\TT^2} + o\left(\frac{1}{\TT^2}\right) \,,
\end{align}
where the omitted sub-leading terms decay as $\exp(-(1 - |u|) \TT)$, and $A_{\mathrm{BG}}(\omega_1,\omega_2)$ is a
$\TT$-independent amplitude which is a rather complicated function of $\omega_1$ and $\omega_2$, that we prefer not to show here. Therefore, in the leading in $\TT$ order and for $\omega_1 \neq \omega_2$ the excess correlation function $J_{\mathrm{BG}}(\omega_1,\omega_2)$ vanishes as a power law,  
in proportion to the second inverse power of the observation time.  We note, as well, that $A_{\mathrm{BG}}(\omega_1,\omega_2)$ vanishes
when either of the frequencies is kept fixed while the other one tends to infinity. In particular, for $\omega_2$ fixed and $\omega_1 \to \infty$, we find
\begin{align}
A_{\mathrm{BG}}(\omega_1,\omega_2) = \frac{4 \left(T_x + u^2 T_y\right)^2 + \omega_2^2 \left(2 T_x + u^2 \left(T_y - T_x\right) 
\right)^2 }{2 \left(1-u^2\right)^2 \left(\omega_2^2 + (1-u)^2\right) \left(\omega_2^2 + (1+u)^2\right)} \frac{1}{\omega_1^2}
 + O\left(\frac{1}{\omega_1^4}\right) \,,
\end{align} 
meaning that $S(\omega_1,\TT)
$ and $S(\omega_2,\TT)$ become statistically independent when either $\TT \to \infty$ or when $|\omega_1 - \omega_2| \to \infty$.
Such a behaviour is identical to the one which we observed for the OU process. Moreover, the covariance function $\overline{b_{\omega_1} b_{\omega_2}}$ has exactly the same form as the one of the OU process, eq.~\eqref{bbOU}.

In ref.~\cite{sara} the joint distribution of the PSD for $X$ and $Y$, for the same values of~$\omega$ and~$\TT$ has been evaluated.

\section{Fractional Brownian motion}
\label{FBM}

We finally consider the fractional Brownian motion (FBM) \cite{ness}. This is a Gaussian process with zero mean and  covariance function
\begin{align}
\label{fBm}
\overline{X(t_1) X(t_2)} = D \left(t_1^{2H} + t_2^{2H} - |t_1 -t_2|^{2H}\right) \,,
\end{align}
where the \textit{Hurst index} $H$ is a real number such that $0 <H < 1$. 
Notice that the FBM is an $H$-parametrized  family of anomalous diffusion processes (except for the case $H = 1/2$,  when one recovers the standard BM with independent increments).
For $H < 1/2$, the process is sub-diffusive as one can readily infer from eq.~\eqref{fBm} by setting $t_1=t_2=t$ to find that the mean-squared displacement obeys $\overline{X^2(t)} = 2 D\, t^{2H}$, where $2H<1$. One can show with little effort that  the increments of the process in this case have negative long-ranged correlations.   
On the contrary, for $H > 1/2$ the exponent $2H$ exceeds unity and one observes a super-diffusive behaviour. In this latter case the increments have positive long-ranged correlations.  
Therefore, one naturally expects for such a family of anomalous diffusion processes a richer behaviour than the one discussed in previous sections.

\subsection{Sub-diffusive fractional Brownian motion}

\begin{figure}
\begin{center}
\includegraphics[width=140mm]{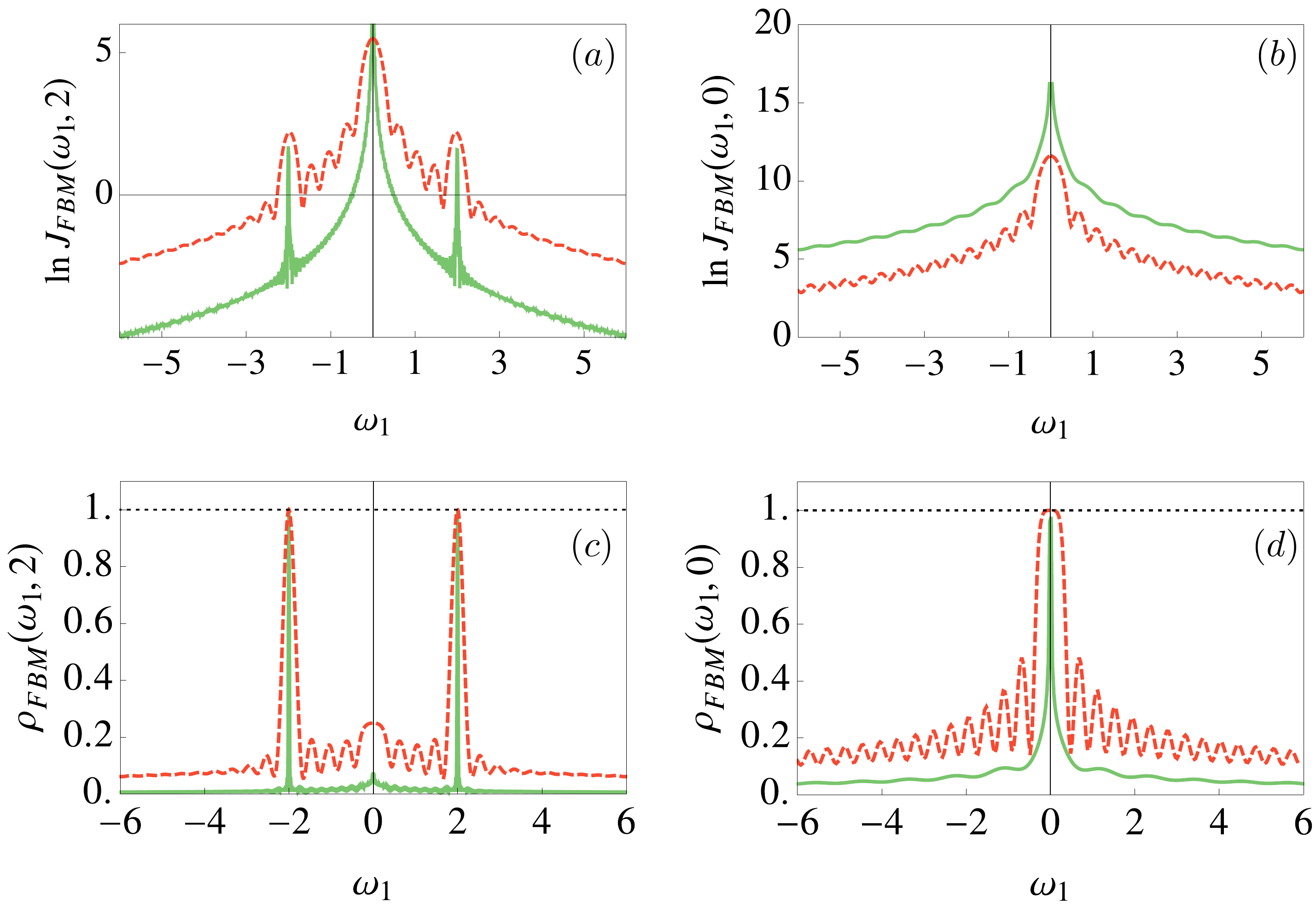}
\end{center}
\caption{Sub-diffusive fractional Brownian motion with $H = 1/4$. Logarithm of the excess correlation function $J_{\mathrm{FBM}}(\omega_1,\omega_2)$ and the Pearson correlation coefficient $\rho_{\mathrm{FBM}}(\omega_1,\omega_2)$ as functions of $\omega_1$ for fixed $\omega_2=2$ (panels (a) and (c)) and for $\omega_2 =0 $ (panels (b) and (d)). Solid lines: $\TT = 200$; dashed lines: $\TT =15$.
}
\label{fig5}
\end{figure}

We start with the case of sub-diffusion, defined by a value of~$H$ satisfying $0 < H < 1/2$. In this case we take advantage of the fact that for $H=1/4$ we can obtain an explicit, albeit lengthy, expression of the excess correlation function  $J_{\mathrm{FBM}}(\omega_1,\omega_2)$. We checked that the behavior for different values of~$H$ in this range, obtained by numerical integration of eq.~\eqref{J}, is qualitatively similar.
Figure~\ref{fig5} shows the excess correlation function  $J_{\mathrm{FBM}}(\omega_1,\omega_2)$ and the Pearson correlation coefficient $\rho_{\mathrm{FBM}}(\omega_1,\omega_2)$ for $H=1/4$ as functions of $\omega_1$ for two fixed values of $\omega_2$: $\omega_2=2$ (panels (a) and (c)) and $\omega_2=0$ (panels (b) and (d)), and for two values of the observation time, namely $\TT = 200$ (solid green curve) and 
 $\TT =15$ (dashed red curve). We observe that $J_{\mathrm{FBM}}(\omega_1,\omega_2)$ exhibits a different behavior from the cases examined so far. Here, in panel (a), the peaks at $\omega_1 = \pm \omega_2$ are much more pronounced than for the BM, and also the central peak at $\omega_1=0$ (and $\omega_2 = 2$) appears to diverge with the observation time, what did not happen for the OU process and for the BG model. Indeed, we find (see~\ref{A}) that for $\omega_1=0$,  and for $(\omega_2 \TT) \to \infty$ (i.e., for either fixed $\omega_2$ and $\TT \to \infty$, or fixed $\TT$ and $\omega_2 \to \infty$),  the excess correlation function $J_{\mathrm{FBM}}(0,\omega)$ obeys
 \begin{align}
 \label{0f}
 J_{\mathrm{FBM}}(0,\omega) \simeq \frac{2 D^2 \TT^{4 H}}{\omega^2} \,.
 \end{align}
 This asymptotic form  holds for arbitrary values of~$H$, i.e., for sub-diffusion, super-diffusion, and also for the Brownian case with $H=1/2$ (see eq.~\eqref{22}). Correction terms  to the asymptotic form in eq.~\eqref{0f} are shown in \ref{A}.
   Interestingly, the $\omega^{-2}$ dependence on the frequency $\omega$ in eq.~\eqref{0f} seems to hold for all the processes studied here. The $\TT$-dependence is, of course, quite different for the processes with a bounded variance (the OU process and the BG model) and for the sub-diffusive FBM for which the variance exhibits an unbounded growth. 
  
Next, on panel (b) of Fig.~\eqref{fig5} we present the excess 
correlation function $J_{\mathrm{FBM}}(\omega_1,\omega_2=0)$ for $\omega_2=0$. 
We observe that, qualitatively, the behaviour appears similar to 
the ones encountered in all our above analyses, i.e., $J_{\mathrm{FBM}}(\omega_1,\omega_2=0)$ has a single peak at $\omega_1=0$. The height of this peak is found analytically to be given by
\begin{align}
\label{00}
J_{\mathrm{FBM}}(0,0) = \frac{2 D^2 \TT^{2+4 H}}{(1+H)^2} \,.
\end{align}
This expression is also valid for arbitrary $H$ (see~\cite{krapf2} and~\ref{A}). In particular, 
setting $H=1/2$ (and hence, $D = T$), we recover our previous eq.~\eqref{23}. Since $J_{\mathrm{FBM}}(0,0)$ is the squared area under the $X(t)$ curve divided by~$\TT$, we find quite naturally that it diverges for the FBM as for the BM, at variance with the bounded processes (OU and BG).

\begin{figure}
\begin{center}
\includegraphics[width=155mm]{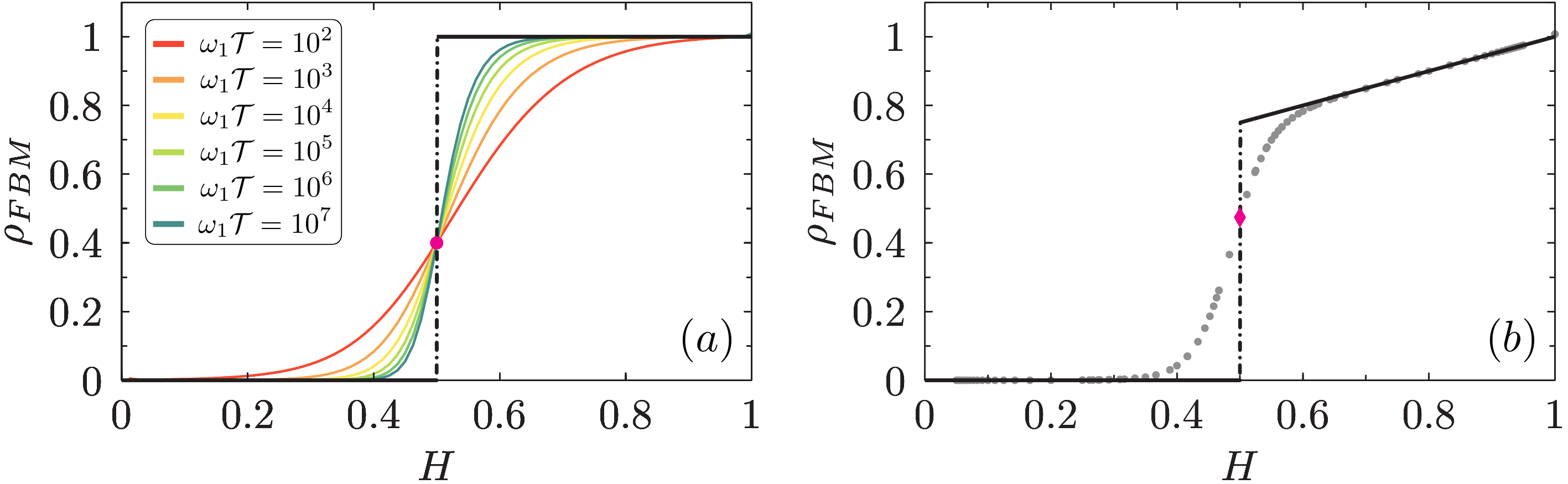}
\end{center}
\caption{Fractional Brownian motion with arbitrary $H \in (0,1)$. Panel (a): Pearson correlation coefficient $\rho_{\mathrm{FBM}}(\omega_1,\omega_2)$ for $\omega_1=1$ and $\omega_2=2$ as function of the Hurst index $H$ for different values of the parameter $\omega_1\TT$ (see the inset). The solid line shows the $\TT\to\infty$ limit, given by $\theta(H-1/2)$, where $\theta(x)$ is the Heaviside function. The value $2/5$ for $H=1/2$ is highlighted. Panel (b): Pearson correlation coefficient $\rho_{\mathrm{FBM}}(\omega_1=0,\omega_2 = 2)$  as function of the Hurst index $H$ for  $\TT = 200$. The solid lines shows the $\TT\to\infty$ limit, given by $\theta(H-1/2)(1+H)/2$ (see eq.~\eqref{hh}). The value $3/(2\sqrt{10})$ (eq.~\eqref{funny}) for $H=1/2$ is highlighted.
}
\label{fig44}
\end{figure}

Conversely, for sub-diffusive FBM the behaviour of the Pearson correlation coefficient is strikingly different from what we have found for the BM and,
in fact, appears to be closer to what we have found for the OU process and for the BG model. Indeed, 
we observe in 
panels (c) and (d) of Fig.~\ref{fig5} that $\rho_{\mathrm{FBM}}(\omega_1,\omega_2)$ for $H=1/4$  
tends to zero as $\TT \to \infty$, both when $\omega_1 \neq \omega_2$ (and both frequencies are not equal to zero) and when $\omega_2 = 0$ and $\omega_1 \neq 0$. This implies, of course, 
that the amplitude $b_\omega$ in eq.~\eqref{prob3} is totally uncorrelated for different values of $\omega$ in this limit. Furthermore, in Fig.~\ref{fig44} we show that this is indeed the case for arbitrary $H$ in the sub-diffusive regime:  $\rho_{\mathrm{FBM}}(\omega_1,\omega_2)$ 
with $\omega_1 \neq \omega$ vanishes as $\TT$ grows for any $H < 1/2$ (see panel (a))  and also $\rho_{\mathrm{FBM}}(\omega_1=0,\omega_2>0)$ is close to zero when  $\TT$ is large (see panel~(b)).

Below we define the rate at which $\rho_{\mathrm{FBM}}(\omega_1,\omega_2)$ vanishes for the sub-diffusive FBM in the limit $\TT \to \infty$ (see \ref{A} for the details of intermediate calculations).
Suppose first that $\omega_1=\omega > 0$ and $\omega_2 = 0$. In this special case the analytical calculations are much simpler  than for arbitrary fixed $\omega_2$.
We recall that $\rho_{\mathrm{FBM}}(\omega,0)$ is formally defined by (see eq.~\eqref{rhor})
\begin{align}
\rho_{\mathrm{FBM}}(\omega,0)= \frac{J_{\mathrm{FBM}}(\omega,0)}{\sqrt{J_{\mathrm{FBM}}(\omega,\omega) J_{\mathrm{FBM}}(0,0)}} \,,
\end{align}
where the large-$\TT$ asymptotic form of $J_{\mathrm{FBM}}(\omega,0)$ is given in eq.~\eqref{0f}, while
$J_{\mathrm{FBM}}(0,0)$ is defined in eq.~\eqref{00}. In turn, we notice that $J_{\mathrm{FBM}}(\omega,\omega)$ is simply the variance of a single-trajectory PSD of a sub-diffusive FBM. This quantity
has been previously determined in~\cite{krapf2} for arbitrary $H \in (0,1)$ and is given by
\begin{align}
\label{V}
J_{\mathrm{FBM}}(\omega,\omega) =  4 D^2 \left(\frac{c_H^2}{\omega^{2+4H}} + \frac{2 c_H}{\omega^{3+2H} \TT^{1-2H}} + \frac{2}{\omega^4 \TT^{2 - 4 H}}\right) \,,
\end{align}
where $c_H = \Gamma(1+2H) \sin(\pi H)$. The expression~\eqref{V} is valid for arbitrary $\omega > 0$ and sufficiently large $\TT$.
Noticing that for $H < 1/2$ the first term in eq.~\eqref{V} defines the dominant large-$\TT$ behaviour and 
combining it with eqs.~\eqref{0f} and \eqref{00}, we therefore conclude that for a sub-diffusive FBM the Pearson correlation coefficient $\rho_{\mathrm{FBM}}(\omega,0)$ vanishes as a power-law when $\TT \to \infty$ as
\begin{align}
\label{vanish}
\rho_{\mathrm{FBM}}(\omega,0) \simeq \frac{1+H}{\sqrt{2} c_H} \frac{1}{\left(\omega \TT\right)^{1-2H}}
\end{align}
Here, two comments are in order. First, as shown in~\ref{A} (see eq.~\eqref{zu4}), the asymptotic form in eq.~\eqref{0f} defines the dominant behaviour in the limit when (sufficiently large) $\TT$ is fixed, while $\omega \to \infty$. Similarly, in this case the dominant large-$\omega$ behaviour of the variance in eq.~\eqref{V} is defined by the first term. As a consequence, the asymptotic form in eq.~\eqref{vanish} describes as well the behaviour of the Pearson coefficient in the limit $\omega \to \infty$ at fixed sufficiently large $\TT$. We therefore conclude that, similarly to the OU process and the BG model, for sub-diffusive FBM, i.e., for any $H \in (0,1/2)$, the correlations between $S(0,\TT)$ and $S(\omega,\TT)$ vanish when either  $\omega$ is fixed and $\TT \to \infty$, or when $\TT$ is fixed and $\omega \to \infty$.
Second, 
 we note that the limit of expression \eqref{vanish} for $H\to 1/2$ yields an incorrect value.
Indeed, for $H=1/2$ the Pearson coefficient in eq.~\eqref{vanish} becomes independent of $\omega$ and $\TT$, but the numerical value is wrong. The point is that 
in order to recover our eq.~\eqref{funny}, we have to take into account all three terms in eq.~\eqref{V}. This implies that the case $H = 1/2$ is singular.

We consider at last the behaviour in a more general case when $\omega_1 \neq \omega_2 > 0$. Relegating the quite tedious calculations to \ref{A}, we find that in the leading order in the limit ${\cal T} \to \infty$, the excess correlation function $J_{\mathrm{FBM}}(\omega_1,\omega_2)$ is given by
\begin{align}
\label{central}
J_{\mathrm{FBM}}(\omega_1,\omega_2) \simeq \frac{8 D^2  {\cal T}^{4H-2}}{\omega_1^2 \omega_2^2} \,,
\end{align}
an expression that holds for any $H \in (0,1)$. Taking into account eq.~\eqref{V}, we obtain
\begin{align}
\rho_{\mathrm{FBM}}(\omega_1,\omega_2) \simeq \frac{2}{c_H^2 (\omega_1 \omega_2)^{1-2H} {\cal T}^{2 - 4H}} \,.
\end{align}
Hence, in this more general case $\rho_{\mathrm{FBM}}(\omega_1,\omega_2)$ also vanishes as ${\cal T} \to \infty$, and the decay is faster than when either of the frequencies is equal to zero (see eq.~\eqref{vanish}).

Summarizing, for a sub-diffusive FBM the two-frequency correlation function in eq.~\eqref{bb} obeys, in the limit $\TT \to \infty$,
\begin{equation}
\label{bbFBMsub}
\overline{b_{\omega_1} b_{\omega_2}} =  
\begin{cases}
1  \, ,     & \text{for } \omega_1 \neq \omega_2; \qquad \omega_1,\omega_2 > 0 ;\\
1  \, ,  & \text{for } \omega_1=0, \, \omega_2 > 0  ;  \\
2 \, , & \text{for }  \omega_1 = \omega_2 \neq 0 ; \\
3 \,, & \text{for }  \omega_1 = \omega_2 = 0 .
\end{cases}
\end{equation}
This implies again that for $\omega_1\neq\omega_2$ the $b_\omega$ are totally uncorrelated.

\subsection{Super-diffusive fractional Brownian motion}

\begin{figure}
\begin{center}
\includegraphics[width=140mm]{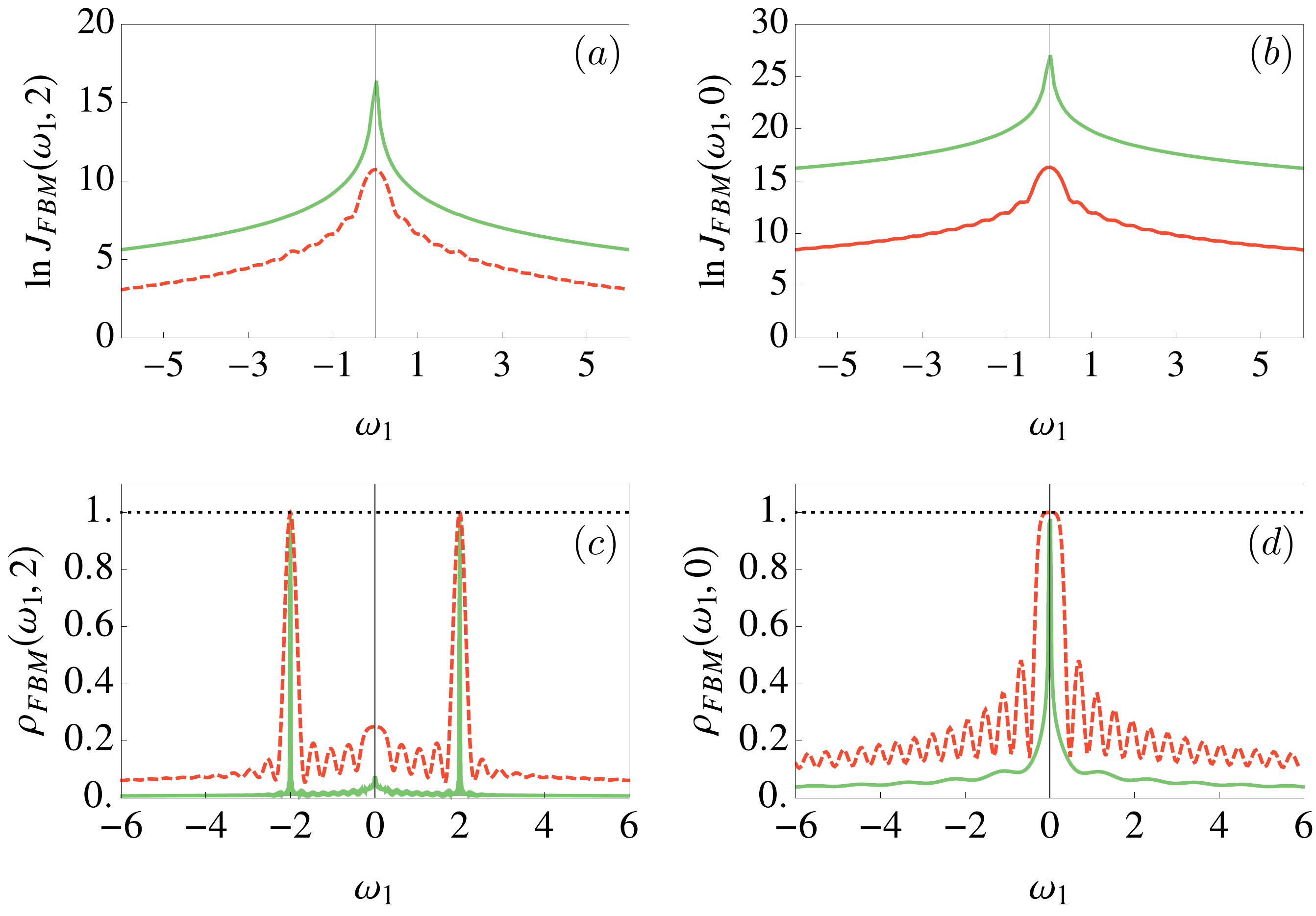}
\end{center}
\caption{Super-diffusive fractional Brownian motion with $H=3/4$. Logarithm  of $J_{\mathrm{FBM}}(\omega_1,\omega_2)$ and the Pearson correlation coefficient $\rho_{\mathrm{FBM}}(\omega_1,\omega_2)$
as functions of $\omega_1$ for fixed $\omega_2=2$ (panels (a) and (c)) and $\omega_2 =0 $ (panels (b) and (d)). Solid lines: $\TT = 200$; dashed lines: $\TT =15$.
}
\label{fig4}
\end{figure}

We turn finally to super-diffusive FBM, i.e.,  to the case with $H > 1/2$. We again exploit the fact that an explicit form of $J_{\mathrm{FBM}}(\omega_1,\omega_2)$ is available for the special value $H=3/4$. In Fig.~\eqref{fig4} we plot $J_{\mathrm{FBM}}(\omega_1,\omega_2)$ for this value of~$H$ as a function of $\omega_1$ for $\omega_2 = 2$ (panel (a)) and $\omega_2=0$ (panel (b)), as well as the corresponding Pearson coefficient $\rho_{\mathrm{FBM}}(\omega_1,\omega_2)$ (panels (c) and (d)). However, we verified that the curves exhibit essentially the same behaviour for any value of $H$ in the super-diffusive regime.   
We find that the behaviour in the super-diffusive case appears to be \textit{qualitatively} different, as compared to the previous situations. We observe in particular that $J(\omega_1,\omega_2=2)$ in
panel (a) of Fig.~\ref{fig4} grows uniformly as $\TT$ grows, while in the previous case it either decreased as $\TT$ increased (for the OU process, the BG model and sub-diffusive FBM), or reached a $\TT$-independent limit, as for the BM (except for the vicinity of $\omega_1 = 0$). Moreover, the peaks at $\omega_1 = \pm 2$ are less pronounced than in the above studied situations and become almost indistinguishable from the base curve for longer observation times. This apparently signals that the correlations between $S(\omega_1,\TT)$ and  $S(\omega_2,\TT)$ become very strong for any different values of $\omega_1$ and $\omega_2$. Of course, this is not totally surprising, because of the strongly correlated increments of the parental process. On the other hand, $J_{\mathrm{FBM}}(\omega_1,0)$ (see panel (b) in Fig.~\ref{fig4}) behaves very similarly to the BM or the sub-diffusive FBM. 

The behaviour of the Pearson correlation coefficient $\rho_{\mathrm{FBM}}(\omega_1,\omega_2=2)$ on panel (c) of Fig.~\ref{fig4} also differs qualitatively from what we have previously observed in two aspects: (i) While there still exist two peaks at $\omega_1 = \pm 2$, there is no peak at $\omega_1 = 0$ and there is instead a dip, and (ii) $\rho_{\mathrm{FBM}}(\omega_1,\omega_2=2)$ raises towards higher values upon an increase of $\cal T$ (except in the vicinity of $\omega_1=0$).  Such an increase with increasing $\TT$ is also apparent in panel (d). The levels which the curves attain for $\TT = 200$ are, however, rather different.

To interpret this behavior, we start by dwelling on $J_{\mathrm{FBM}}(\omega_1,\omega_2=0)$, which defines the value of the excess correlation function at the location of the dip on panel (c) and on its behavior depicted on the panel (d). We focus first on the limit $\TT \to \infty$ with $\omega_1$ kept fixed. In this limit, the dominant contribution to $J_{\mathrm{FBM}}(\omega,\omega)$  in eq.~\eqref{V} is provided by the third term in the brackets, while the first sub-dominant correction is given by the second term. The first term is irrelevant, unlike for the case of sub-diffusion, for which it plays the dominant role. Recalling the expressions \eqref{0f} and~\eqref{00}, we find that the Pearson correlation coefficient admits the following asymptotic form for large~$\TT$:
 \begin{align}
 \label{hh}
 \rho_{\mathrm{FBM}}(\omega_1,0) = \frac{H+1}{2}\left(1 - \frac{c_H}{2 (\omega_1 \TT)^{2H-1}} + \mathrm{o}\left(\frac{1}{\TT^{2H-1}}\right) \right) \,.
 \end{align}  
Therefore, $\rho_{\mathrm{FBM}}(\omega_1,0)$ indeed increases as $\TT$ grows, saturating at the constant value $(H+1)/2$. Such a behaviour is clearly seen on panel (b) of Fig. \ref{fig44}, in which we plot $\rho_{\mathrm{FBM}}(0,\omega_2=2)$ as function of $H$ for a sufficiently large observation time ($\TT = 200$), as well as the leading term in eq.~\eqref{hh}. We note that expression \eqref{hh} yields an incorrect value in the limit $H \to 1/2$. We thus have a discontinuity at $H = 1/2$:  a drop from the value $3/4$ predicted by eq.~\eqref{hh} when we extrapolate $H$ to $1/2$, to the smaller actual value $ 3/(2 \sqrt{10}) \approx 0.474 $ (see eq.~\eqref{funny}).

Next, we look at the correlations of $S(\omega_1,{\cal T})$ and $S(0,{\cal T})$, in the limit $\omega_1\to\infty$ with ${\cal T}$ kept fixed and sufficiently large to ensure the validity of eq. \eqref{V}. Inspecting eq. \eqref{V}, we obtain that the leading behaviour in this limit is provided by the third term in the brackets and the first sub-dominant correction comes from the second term. This straightforwardly implies that  $ \rho_{\mathrm{FBM}}(\omega_1,0)$ in this limit obeys eq.~\eqref{hh}.
Thus the Pearson correlation coefficient saturates at the constant value $(1+H)/2$ and correlations do not decouple completely. Such a behaviour is apparent in Fig.~\ref{fig44}~(panel (b)).

Lastly, we turn to the more general case when $\omega_1 \neq \omega_2$,  both non-zero. Taking advantage of  eqs.~\eqref{V} and \eqref{central}, we arrive at
\begin{align}
\lim_{\TT\to\infty}\rho_{\mathrm{FBM}}(\omega_1,\omega_2) = 1.
\end{align}
This means that for a super-diffusive FBM the variables $S(\omega_1,{\cal T})$ and $S(\omega_2,{\cal T})$ become \textit{completely correlated} in the limit ${\cal T} \to \infty$.

Summarizing, we have, in the limit $\TT \to \infty$,
\begin{equation}
\label{bbFBMsup}
\overline{b_{\omega_1} b_{\omega_2}} =  
\begin{cases}
3  \, ,     & \text{for }\omega_1 \neq \omega_2; \qquad \omega_1,\omega_2 > 0; \\
2 + H \, ,  &\text{for }\omega_1=0, \, \omega_2 > 0;   \\
3\, , & \text{for } \omega_1=\omega_2 \neq 0; \\
3 \,, & \text{for }\omega_1 = \omega_2 = 0 .
\end{cases}
\end{equation}
We thus have complete correlations among the $b_\omega$, when $\omega\neq 0$, and a partial one when one of the $\omega$ vanishes. We can therefore conclude that the correlations of $b_\omega$ are a powerful probe to distinguish between the sub-diffusive, diffusive and super-diffusive cases of the BM.

\section{Conclusions}
\label{conc2}

In conclusion, we examined the correlations of the reduced spectral density $b_\omega = S(\omega,\TT)/\mu(\omega,\TT)$ for a class of Gaussian random processes, where $S(\omega,\TT)$ is the spectral density of a single trajectory $X(t)$ of the process, eq.~\eqref{def2}, $\omega$ is the frequency, $\TT$ is the observation time, and $\mu(\omega,\TT)=\overline{S(\omega,\TT)}$ is the average of $S(\omega,\TT)$ over all realizations of the process. It was shown in previous works that for a number of different classes of Gaussian stochastic processes, the spectral density $S(\omega,\TT)$ is a random variable whose probability distribution depends on $S(\omega,\TT)$ and $\mu(\omega,\TT)$ only in the combination $b_\omega=S(\omega,\TT)/\mu(\omega,\TT)$. (The resulting distribution depends moreover on $\omega$ only via the coefficient of variation $\gamma(\omega,\TT)$.) We have therefore looked at $b_\omega$ as an $\omega$-dependent random quantity. Looking at the behavior of this quantity for different values of $\omega$ and for different cases, we have shown that the values of $b_\omega$ can be either fully uncorrelated (e.g., for the Ornstein-Uhlenbeck process, for the Brownian gyrator, and for the subdiffusive Brownian motion), partially correlated (e.g., for the ordinary, diffusive Brownian motion) or totally correlated (e.g., for superdiffusive Brownian motion, when both frequencies are non-vanishing). Therefore $b_\omega$ can be used as a sensitive probe of the nature of the process, that is available upon inspection of a single trajectory, provided the observation time $\TT$ is large enough. It can thus supplement in this role the coefficient of variation $\gamma$. Indeed, it was suggested in~\cite{krapf2} to use $\gamma$ as a criterion that permits to distinguish between different kinds of Gaussian processes. This requires, however, to evaluate either the large-$\omega$ asymptotical behaviour of $\gamma$ in case of sub-diffusion and diffusion, or its ageing properties (i.e.,its $\TT$-dependence) for superdiffusion, what may be somewhat difficult to access.  Thus using $b_\omega$  appears advantageous, because it does not require to evaluate the asymptotical behavior, and the values of the frequencies can be arbitrarily chosen. In particular, if we look at the $\overline{b_{\omega}b_{\omega'}}$ correlation function when one of the frequencies vanish in the super-diffusive FBM case, we are able to extract the value of the Hurst parameter~$H$.

It remains the challenging problem of extending this kind of analysis to non-Gaussian processes.

\appendix

\section{Analytical results for FBM.}
\label{A}

\subsection{Excess correlation function $J_{\mathrm{FBM}}(0,\omega)$.}

We take advantage of the representation of the
covariance function of the FBM with arbitrary
$H$ in the integral form (see~\cite{kacha}), in which the kernel
has a convenient factorised dependence on $t_1$ and $t_2$ :
\begin{equation}
\begin{split}
 \label{cof}
 t_1^{2H} + t_2^{2H} - |t_1-t_2|^{2H} &= \frac{2 c_H}{\pi} \int^{\infty}_0 \frac{\D z}{z^{2H+1}}\\
 &\qquad {}\times  \left[ \sin(z t_1) \sin(z t_2)  + (1- \cos(z t_1))(1-\cos(z t_2)) \right] \,,
\end{split}
\end{equation}
where 
\begin{equation}\label{eq:cH}
c_H = \Gamma(2H+1) \sin(\pi H).
\end{equation}
Inserting this expression into the definition of $J(\omega_1,\omega_2)$, we perform the integrations over the time variables, and eventually, over $z$. In doing so, we find that $J(0,\omega)$ is given by
\begin{equation}
\label{zu}
J_{\mathrm{FBM}}(0,\omega) = D^{2} \TT^{2+4H} \mathcal{J}_{H}(w) \, , \quad w= \omega \TT \,,
\end{equation}
where $\mathcal{J}_{H}(w)$ is the following function of the dimensionless variable $w$ only:
\begin{equation}
\label{ }
\mathcal{J}_{H}(w) = \frac{1}{2 (1+H)^{2} (1+2H)^{4}} \left[ \frac{w^2}{(2H+3)^{2}} \mathcal{K}_{1}^{2}(w,H) + \mathcal{K}_{2}^{2}(w,H) \right]\,.
\end{equation}
In this expression, one has
\begin{equation}
\begin{split}
\mathcal{K}_{1}(w,H) & =(2 H+3)  \, _1F_2\left(1;H+\frac{3}{2},H+2;-\frac{w^2}{4}\right) \\
&\qquad {} -2  \, _1F_2\left(2;H+2,H+\frac{5}{2};-\frac{w^2}{4}\right) \\ 
&\qquad {} - (1+2H) \left(4 H^2+8 H+3\right)  \, _1F_2\left(H+1;\frac{3}{2},H+2;-\frac{w^2}{4}\right) \\ 
&\qquad {} + 2 (1+H) (1+2H)  {}_{1}F_2\left(H+\frac{3}{2};\frac{3}{2},H+\frac{5}{2};-\frac{w^2}{4}\right) \\
&\qquad {} - 2 (1+H) (1+2H) (2 H+3) \frac{(1-\cos (w))}{w^2} \, ,
\end{split}
\end{equation}
and
\begin{equation}
\begin{split}
\mathcal{K}_{2}(w,H) & = -2 (H+1)   \, _1F_2\left(1;H+1,H+\frac{3}{2};-\frac{w^2}{4}\right)  
\\
&\qquad {} -  \, _1F_2\left(H+1;\frac{1}{2},H+2;-\frac{w^2}{4}\right) \\ 
&\qquad {} -2 H   \, _1F_2\left(H+1;\frac{1}{2},H+2;-\frac{w^2}{4}\right) \\
&\qquad {}+  \, _2F_3\left(1,\frac{3}{2};\frac{1}{2},H+\frac{3}{2},H+2;-\frac{w^2}{4}\right) \\
&\qquad {}+2 \left(2 H^2+3 H+1\right)   \, _1F_2\left(H+\frac{1}{2};\frac{1}{2},H+\frac{3}{2};-\frac{w^2}{4}\right) \\ 
&\qquad {} + 2 (H^2 + 3 H + 1) \frac{\sin(w)}{w} \,,
\end{split}
\end{equation}
where $_1F_2$ and $_2F_3$ are hypergeometric functions. The expression \eqref{zu} is valid for arbitrary 
$H \in (0,1)$ and for arbitrary values of~$\TT$ and~$\omega$. In particular, for $\omega = 0$ it reduces to expression \eqref{00}.

We next consider the large-$w$ behaviour of $J_{\mathrm{FBM}}(0,\omega)$ in eq.~\eqref{zu}, which is realised when either $\omega$ is fixed and $\TT$ tends to infinity, or alternatively, $\TT$ is fixed but $\omega \to \infty$. Expanding the hypergeometric functions in the limit when $w=(\omega\TT) \to \infty$, we find 
\begin{equation}
\begin{split}
\label{zu4}
&J_{\mathrm{FBM}}(0,\omega) = \frac{2 D^2 \TT^{4 H}}{\omega^2}\left[ \Bigg(\sin(w) -\frac{c_H}{w^{2H}} - \frac{1+(1-2H) \cos(w)}{w} +\mathcal{O}\left(\frac{1}{w^{2H+1}}\right)\Bigg)^2\right. \\
&\qquad\qquad \left.{}+ 
 \Bigg(\cos(w) -\frac{c_H \cot(w)}{w^{2H}} + \frac{(1-2H) \sin(w)}{w} +\mathcal{O}\left(\frac{1}{w^{2H+1}}\right)\Bigg)^2\right],
\end{split}
\end{equation}
where $c_H$ is defined in eq.~\eqref{eq:cH}. This expression is dominated by the first terms in the parentheses, that oscillate as functions of~$w$. We thus arrive at the asymptotic form in eq.~\eqref{0f}. Notice that 
this expression is valid
for any $H$ such that $0< H< 1$. The finite-$w$ corrections to the leading behaviour in eq.~\eqref{0f} are however different in the cases $H < 1/2$, $H = 1/2$ or $H > 1/2$. Indeed, in case of sub-diffusion, the dominant correction terms is given by the second terms in brackets 
in eq.~\eqref{zu4}, while in the super-diffusive case is given by the third terms.  

\subsection{Large-$\TT$ asymptotic behaviour of the excess correlation function $J_{\mathrm{FBM}}(\omega_1,\omega_2)$.}

Here we discuss the large-$\TT$ behaviour of the excess correlation function $J_{\mathrm{FBM}}(\omega_1,\omega_2)$ in the more general case when $\omega_1 \neq \omega_2$ and both frequencies are non-vanishing.  We start by appropriately rescaling the integration variables in eqs.~\eqref{W} to get, e.g., for $W_{\cc}$, the following representation
\begin{equation}
\label{met}
\frac{W_{\cc}(\omega_1,\omega_2;\TT)}{D \TT^{2H+1} } = \int^1_0 \D\tau_1\; \cos(\omega_1 \TT \tau_1) \int^1_0 \D\tau_2\; \left(\tau_1^{2H} + \tau_2^{2H} - |\tau_1 - \tau_2|^{2H}\right)  \cos(\omega_2 \TT \tau_2) \,.
\end{equation}
Similar expression can be obtained for $W_{\cs}$, $W_{\sc}$ and $W_{\ss}$ and will differ from the one in eq.~\eqref{met} only by the sine and cosine factors. 
The integrals corresponding to the first and the second terms in the covariance function can be performed analytically 
and give, in the leading in the limit order for $\TT \to \infty$,
\begin{equation}\label{simple}
\begin{split}
&\int^1_0 \D\tau_1\; \left\{\begin{array}{lr}
        \cos(\omega_1 \TT \tau_1) \\
        \sin(\omega_1 \TT \tau_1)\\
        \end{array}\right\} 
        \int^1_0 \D\tau_2\; \tau_2^{2H} \left\{\begin{array}{lr}
        \cos(\omega_2 \TT \tau_2) \\
        \sin(\omega_2 \TT \tau_2)\\
        \end{array}\right\} \\
        &\qquad = \left\{\begin{array}{lr}
        \sin(\omega_1 \TT) \\
        1-\cos(\omega_1 \TT)\\
        \end{array}\right\} 
         \left\{\begin{array}{lr}
        \sin(\omega_2 \TT) \\
        -\cos(\omega_2 \TT)\\
        \end{array}\right\} \frac{1}{\omega_1 \omega_2 \TT^2}  + \mathcal{O}\left(\frac{1}{\TT^3}\right)\,.
\end{split}
\end{equation}
This asymptotic expression  is valid for any $0< H < 1$. The analysis of the integrals involving the third, non-local term in the covariance of the FBM is a bit more delicate and we first would like to cast it in a more transparent form. To this end, we expand both cosine functions in powers of $\tau_1$ and $\tau_2$, perform the integrals, and then resum the series. In doing so, we get
\begin{equation}
\begin{split}
&\int^1_0 \D\tau_1\; \cos(\omega_1 \TT \tau_1) \int^1_0 \D\tau_2 \;  |\tau_1 - \tau_2|^{2H}  \cos(\omega_2 \TT \tau_2)  \\
&\qquad =  \frac{\Gamma(2H+1)}{(\omega_2 \TT)^{2H+2}} \int^{\omega_2 \TT}_0 u^{2H+1} du \, \cos\left(\frac{\omega_1}{\omega_2} u\right) \, E_{2,2(H+1)}(-u^2) \\
&\qquad\qquad {}+ \frac{\Gamma(2H+1)}{(\omega_1 \TT)^{2H+2}} \int^{\omega_1 \TT}_0 u^{2H+1} du \,\cos\left(\frac{\omega_2}{\omega_1} u\right) \, E_{2,2(H+1)}(-u^2) \,,
\end{split} 
\end{equation}
where
\begin{align}
E_{2,2(H+1)}(u) = \sum_{k=0}^{\infty} \frac{u^k}{\Gamma(2 k + 2(H+1))}
\end{align}
is the Mittag-Leffler function, whose behaviour is well-documented \cite{mittag}. In a similar way, we obtain 
the following results:
\begin{equation}\label{eq:primo}
\begin{split}
&\int^1_0 \D\tau_1\; \cos(\omega_1 \TT \tau_1) \int^1_0 \D\tau_2 \;  |\tau_1 - \tau_2|^{2H}  \sin(\omega_2 \TT \tau_2)  \\
&\qquad =  \frac{\Gamma(2H+1)}{(\omega_2 \TT)^{2H+2}} \int^{\omega_2 \TT}_0 u^{2H+2} \D u \; \cos\left(\frac{\omega_1}{\omega_2} u\right) \, E_{2,2H+3}(-u^2)  \,;
\end{split} 
\end{equation}
\begin{equation}\label{eq:secondo}
\begin{split}
&\int^1_0 \D\tau_1\; \sin(\omega_1 \TT \tau_1) \int^1_0 d\tau_2 \,  |\tau_1 - \tau_2|^{2H}  \cos(\omega_2 \TT \tau_2)  \\
&\qquad =  \frac{\Gamma(2H+1)}{(\omega_1 \TT)^{2H+2}} \int^{\omega_1 \TT}_0 u^{2H+2} \D u \; \cos\left(\frac{\omega_2}{\omega_1} u\right) \, E_{2,2H+3}(-u^2)  \,;
\end{split} 
\end{equation} 
\begin{equation}\label{eq:terzo}
\begin{split}
&\int^1_0 \D\tau_1\; \sin(\omega_1 \TT \tau_1) \int^1_0 \D\tau_2 \;  |\tau_1 - \tau_2|^{2H}  \sin(\omega_2 \TT \tau_2)  \\
&\qquad =  \frac{\Gamma(2H+1)}{(\omega_1 \TT)^{2H+2}} \int^{\omega_1 \TT}_0 u^{2H+2} \D u \;
\sin\left(\frac{\omega_2}{\omega_1} u\right) \, E_{2,2H+3}(-u^2) \\
&\qquad\qquad {} + \frac{\Gamma(2H+1)}{(\omega_2 \TT)^{2H+2}} \int^{\omega_2 \TT}_0 u^{2H+2} \D u\; \sin\left(\frac{\omega_1}{\omega_2} u\right) \, E_{2,2H+3}(-u^2)  \,.
\end{split} 
\end{equation} 

We now turn to the analysis of the  leading large-${\cal T}$ asymptotic behaviour of expressions \eqref{eq:primo} to~\eqref{eq:terzo}. To this end, we first note that the leading large-$u$ asymptotic behaviour of the Mittag-Leffler function obeys
\begin{align}
E_{2,\beta}(-u^2) \simeq \frac{1}{\Gamma(\beta - 2)} \frac{1}{u^2} \,.
\end{align}
Consequently, the integral which enters eq.~\eqref{eq:primo} diverges on the upper terminal of 
integration in the super-diffusive case, and converges to a finite value when ${\cal T} \to \infty$ in the sub-diffusive case. This yields
\begin{equation}
\begin{split}
\label{intcc}
&\int^1_0 \D\tau_1 \cos(\omega_1 {\cal T} \tau_1) \int^1_0 \D\tau_2 \;  |\tau_1 - \tau_2|^{2H}  \cos(\omega_2 {\cal T} \tau_2) \\ 
& \qquad \simeq \begin{cases}
\dfrac{2H}{\omega_1 \omega_2 {\cal T}^3} \left(\dfrac{\sin(\omega_1 {\cal T})}{\omega_2} + \dfrac{\sin(\omega_2 {\cal T})}{\omega_1}\right)\,, &\text{for } H \in (1/2,1),\\
\dfrac{\cos(\pi H) \Gamma(2H+1)\left(\omega_2^{2H} - \omega_1^{2H}\right)}{\omega_1^{2H} \omega_2^{2H} \left(\omega_2^2 - \omega_1^{2}\right) {\cal T}^{2H+2}} \,, & \text{for } H \in (0,1/2) \,,
\end{cases}
\end{split}
\end{equation}
where we took advantage of the identity
\begin{align}
\int^{\infty}_0 u^{2H+1}\, \D u \;
\cos(\lambda u)
\, E_{2,2(H+1)}(-u^2) = \frac{\cos(\pi H)}{|\lambda|^{2H} \left(1 - \lambda^2\right)},
 \end{align}
which holds for $H \in (0,1/2)$. Comparing eq.~\eqref{eq:primo} and eq.~\eqref{simple}, we infer that the dominant contribution to the large-${\cal T}$ behaviour of $W_{\cc}(\omega_1,\omega_2;{\cal T})$ is provided by the latter, such that $W_{\cc}$ has the following asymptotic form
\begin{align}
W_{\cc}(\omega_1,\omega_2;{\cal T}) \simeq \frac{2 D \sin(\omega_1 {\cal T}) \sin(\omega_2 {\cal T})}{\omega_1 \omega_2 } {\cal T}^{2H-1} \,,
\end{align}
which is valid for any $H$.

Further on, inspecting the kernels in the integrals in eqs.~\eqref{eq:primo}, \eqref{eq:secondo} and~\eqref{eq:terzo}, we realise that the large-$u$ tails do not decay fast enough to ensure the convergence of the integrals even in the sub-diffusive case.  Performing some rather straightforward calculations, we then find that for any $H \in (0,1)$ the leading asymptotic behaviour of the integrals is given by
\begin{equation}
\begin{split}
&\int^1_0 \D\tau_1\; \cos(\omega_1 {\cal T} \tau_1) \int^1_0 \D\tau_2 \;  |\tau_1 - \tau_2|^{2H}  \sin(\omega_2 {\cal T} \tau_2) \simeq \frac{\sin(\omega_1 {\cal T})}{\omega_1 \omega_2 {\cal T}^2} \,, \\
&\int^1_0 \D\tau_1 \;\sin(\omega_1 {\cal T} \tau_1) \int^1_0 \D\tau_2 \;  |\tau_1 - \tau_2|^{2H}  \cos(\omega_2 {\cal T} \tau_2) \simeq \frac{\sin(\omega_2 {\cal T})}{\omega_1 \omega_2 {\cal T}^2} \,, \\
\end{split}  
\end{equation}
and
\begin{equation}
\begin{split}
&\int^1_0 \D\tau_1\; \sin(\omega_1 {\cal T} \tau_1) \int^1_0 \D\tau_2 \;  |\tau_1 - \tau_2|^{2H}  \sin(\omega_2 {\cal T} \tau_2)  \simeq - \frac{\cos(\omega_1 {\cal T}) + \cos(\omega_2 {\cal T})}{\omega_1 \omega_2 {\cal T}^2} \,.
\end{split}  
\end{equation}
Taking into account the expressions \eqref{simple}, we eventually find
\begin{equation}
\begin{split}
W_{\cs}(\omega_1,\omega_2;{\cal T})  &\simeq - \frac{2 D \sin(\omega_1 {\cal T}) \cos(\omega_2 {\cal T})}{\omega_1 \omega_2 } {\cal T}^{2H-1}  \,, \\
W_{\sc}(\omega_1,\omega_2;{\cal T})  &\simeq - \frac{2 D \cos(\omega_1 {\cal T}) \sin(\omega_2 {\cal T})}{\omega_1 \omega_2 } {\cal T}^{2H-1}  \,,
\end{split}  
\end{equation}
and
\begin{align}
W_{ss}(\omega_1,\omega_2;{\cal T})  &\simeq \frac{2 D \cos(\omega_1 {\cal T}) \cos(\omega_2 {\cal T})}{\omega_1 \omega_2 } {\cal T}^{2H-1}  \,,
\end{align}
which yields our asymptotic expression for $J_{\mathrm{FBM}}(\omega_1,\omega_2)$, eq.~\eqref{central}.


\section*{Acknowledgments}

The authors wish to thank Eli Barkai and Sergio Ciliberto for many helpful discussions. AS and LP acknowledge the warm hospitality of LPTMC, Sorbonne Universit\'e, respectively in October-December 2021 and in April 2022. AS acknowledges FWF Der Wissenschaftsfonds for funding through the Lise-Meitner Fellowship (Grant No.\ M 3300-N). LR acknowledges the support of Italian National Group of Mathematical Physics (GNFM) of INDAM, and of Ministero dell’Istruzione e dell'Università e della Ricerca (MIUR), Italy, Grant No.\ E11G18000350001 ``Dipartimenti di Eccellenza 2018–2022''.

\end{document}